\documentclass[11pt]{article}
\usepackage{times}
\usepackage{epsfig}

\makeatletter

\makeatletter

\DeclareSymbolFont{boldmath}{OML}{cmm}{b}{it}
\DeclareSymbolFontAlphabet{\mathb}{boldmath}
\DeclareMathAlphabet{\Bbb}{U}{msb}{m}{n}
\DeclareMathAlphabet{\mathfrak}{U}{euf}{b}{n}

\DeclareMathSymbol{\balpha}{0}{boldmath}{"0B}
\DeclareMathSymbol{\bbeta}{0}{boldmath}{"0C}
\DeclareMathSymbol{\bgamma}{0}{boldmath}{"0D}
\DeclareMathSymbol{\bdelta}{0}{boldmath}{"0E}
\DeclareMathSymbol{\bepsilon}{0}{boldmath}{"0F}
\DeclareMathSymbol{\bzeta}{0}{boldmath}{"10}
\DeclareMathSymbol{\bfeta}{0}{boldmath}{"11}
\DeclareMathSymbol{\btheta}{0}{boldmath}{"12}
\DeclareMathSymbol{\biota}{0}{boldmath}{"13}
\DeclareMathSymbol{\bkappa}{0}{boldmath}{"14}
\DeclareMathSymbol{\blambda}{0}{boldmath}{"15}
\DeclareMathSymbol{\bmu}{0}{boldmath}{"16}
\DeclareMathSymbol{\bnu}{0}{boldmath}{"17}
\DeclareMathSymbol{\bxi}{0}{boldmath}{"18}
\DeclareMathSymbol{\bpi}{0}{boldmath}{"19}
\DeclareMathSymbol{\brho}{0}{boldmath}{"1A}
\DeclareMathSymbol{\bsigma}{0}{boldmath}{"1B}
\DeclareMathSymbol{\btau}{0}{boldmath}{"1C}
\DeclareMathSymbol{\bupsilon}{0}{boldmath}{"1D}
\DeclareMathSymbol{\bphi}{0}{boldmath}{"1E}
\DeclareMathSymbol{\bchi}{0}{boldmath}{"1F}
\DeclareMathSymbol{\bpsi}{0}{boldmath}{"20}
\DeclareMathSymbol{\bomega}{0}{boldmath}{"21}
\DeclareMathSymbol{\beps}{0}{boldmath}{"22}
\DeclareMathSymbol{\bthet}{0}{boldmath}{"23}
\DeclareMathSymbol{\bomeg}{0}{boldmath}{"24}
\DeclareMathSymbol{\bvphi}{0}{boldmath}{"27}

\@addtoreset{equation}{section}
\newcommand{\nwl}{\nonumber\\}
\newcommand{\txt}[1]{\quad\mathrm{#1}\quad}

\newcommand{\sref}[1]{section~\ref{#1}}

\newcommand{\fref}[1]{figure~\ref{#1}}
\newcommand{\Fref}[1]{Figure~\ref{#1}}
\newcommand{\eref}[1]{(\ref{#1})}
\newcommand{\follows}{\quad\Rightarrow\quad}
\newcommand{\equivalent}{\quad\Leftrightarrow\quad}
\newcommand{\ft}[2]{{\textstyle{{#1}\over{#2}}}}
\newcommand{\del}{\partial} 
\newcommand{\dd}{\mathrm{d}} 

\newcommand{\Tr}{\mathop{\mathrm{Tr}}}
\newcommand{\Trr}[1]{\Tr(#1)}
\newcommand{\inv}{\!^{-1}}
\newcommand{\expo}[1]{e^{#1}}

\newcommand{\grp}{\mathsf{SL}(2)}
\newcommand{\alg}{\mathfrak{sl}(2)}

\newcommand{\ZZ}{\Bbb{Z}}
\newcommand{\RR}{\Bbb{R}}

\newcommand{\ads}{\mathcal{S}}
\newcommand{\scri}{\mathcal{J}}

\newcommand{\bnd}{\Lambda}   
\newcommand{\ext}{\Sigma}  
\newcommand{\wrm}{\Omega} 
\newcommand{\sng}{\Gamma} 
\newcommand{\tor}{\Pi}   
\newcommand{\ctc}{\Delta}  
\newcommand{\cov}[1]{\widetilde #1}
\newcommand{\crv}{\lambda}  
\newcommand{\jmp}{\beta}  

\newcommand{\one}{\mathbf{1}}
\newcommand{\gam}{\bgamma}
\newcommand{\xx}{\mathb{x}}
\newcommand{\yy}{\mathb{y}}
\newcommand{\zz}{\mathb{z}}
\newcommand{\uu}{\mathb{u}}
\newcommand{\vv}{\mathb{v}}
\newcommand{\mm}{\mathb{m}}
\newcommand{\nn}{\mathb{n}}
\newcommand{\gx}{\mathb{g}}
\newcommand{\hx}{\mathb{h}}

\newcommand{\ch}{\chi}
\newcommand{\ph}{\varphi}
\newcommand{\qh}{\theta}

\newcommand{\dis}{\delta}   
\newcommand{\erg}{\epsilon}
\newcommand{\prt}{p}   
\newcommand{\srf}{s}  
\newcommand{\hor}{h}  
\newcommand{\chr}{c} 
\newcommand{\iso}{f}   
\newcommand{\riso}{g}  
\newcommand{\rot}{r}  
\newcommand{\id}{\mathrm{id}}

\newcommand{\kll}{\xi}  
\newcommand{\kllr}{\xi_{\mathrm{rot}}}   
\newcommand{\kllt}{\xi_{\mathrm{time}}}  
\newcommand{\kllh}{\xi_{\mathrm{hor}}}   
\newcommand{\len}{\ell}    
\newcommand{\ang}{\omega} 

\makeatother

\begin{document}

\begin{flushright}
MZ-TH/99-17\\
gr-qc/9905030
\end{flushright}

\begin{center}
  \LARGE \textbf{The Anti-de~Sitter Gott Universe:\\[1ex]
    A Rotating BTZ Wormhole}
\end{center}
 
\vspace*{8mm}

\begin{center}
  \textbf{S\"oren Holst}\\[2ex]
  Fysikum, Stockholms Universitet\\[1ex]
  Vanadisv\"agen 9, Box 6730, S-11385 Stockholm\\[1ex]
  E-mail: holst@physto.se\\[2ex]
  and\\[2ex]
  \textbf{Hans-J\"urgen Matschull}\\[2ex]
  Institut f\"ur Physik, Johannes Gutenberg-Universit\"at\\[1ex]
  Staudingerweg 7, D-55099 Mainz\\[1ex]
  E-mail: matschul@thep.physik.uni-mainz.de
\end{center}

\vspace*{10mm}

\begin{abstract}
  Recently it has been shown that a 2+1 dimensional black hole can be
  created by a collapse of two colliding massless particles in otherwise
  empty anti-de~Sitter space. Here we generalize this construction to
  the case of a non-zero impact parameter. The resulting spacetime,
  which may be regarded as a Gott universe in anti-de~Sitter background,
  contains closed timelike curves. By treating these as singular we are
  able to interpret our solution as a rotating black hole, hence
  providing a link between the Gott universe and the BTZ black hole.
  When analyzing the spacetime we see how the full causal structure of
  the interior can be almost completely inferred just from
  considerations of the conformal boundary.
\end{abstract}

\newpage

\section*{Outline}
The Gott universe \cite{Gott} and the BTZ black hole \cite{BTZ,BHTZ} are
two well known solutions to Einstein's equations in 2+1 dimensions, the
latter involving a cosmological constant.  Both can be constructed by
identifying points in some originally maximally symmetric spacetime.
Thus, starting from three dimensional Minkowski space, cutting out two
wedges in an appropriate way and identifying their faces, yields the
Gott universe. It can be regarded as a spacetime containing two point
particles passing each other at a finite distance.  What makes this
spacetime remarkable is that when the particles' energy is sufficiently
large, then it contains closed timelike curves.  Hence, it is often
referred to as the Gott time machine.

The BTZ solution on the other hand is constructed from anti-de~Sitter
space. It may be thought of as the negatively curved analogue to the
Misner universe \cite{Misner}, or Grant space \cite{Grant}. It is
obtained from anti-de~Sitter space by taking the quotient with respect
to a discrete isometry group generated by a single element. The
resulting spacetime is interesting since, contrary to its flat space
cousins, it may be interpreted as a black hole. This is due to the
exotic structure of the anti-de~Sitter infinity. In the sense of
conformal compactifications, spacelike and lightlike infinity of
anti-de~Sitter space coincide. They form the surface of the cylinder,
which has its own causal structure.  This is very different from the
situation in Minkowski space.  However, in order for the black hole
interpretation to go through, one must agree to regard the subset of the
spacetime that is filled with closed timelike curves, which result from
taking the quotient, as singular.

In this paper we want to present a solution that reveals the close
relation between these two spacetimes, the Gott time machine and the BTZ
black hole. Essentially, what we are going to do is to perform the Gott
construction in anti-de~Sitter space instead of Minkowski space. In
order to simplify the discussion we will take our particles to be
massless, moving on lightlike geodesics. We then treat the resulting
spacetime from the BTZ point of view. That is, instead of considering it
to be a time machine, we declare the subset containing the closed
timelike curves to be singular. This enables us to interpret our
solution as a black hole with horizons. Since it is locally isometric to
anti-de~Sitter space, its exterior region will be that of a BTZ black
hole.

From this point of view, our solution also furnishes the rotating
generalization of the black hole formed by colliding point particles
\cite{Matschull}. Together with earlier results \cite{Mann}, this
implies that the three dimensional BTZ black holes not only share many
important properties with ordinary, four dimensional black holes. They
can also be formed by a matter collapse. The advantage of the three
dimensional toy model with point particles is that the collapse can be
described as an exact solutions to Einstein's equation, with the matter
being reduced to a very simple, and in fact minimal system, just
consisting of two massless point particles. This could finally allow a
fully dynamical description, in which the physical degrees of freedom of
the particles are treated as variables, and of which the spacetime we
are going to construct is just a special, though generic, solution.

Let us give a brief summary of the construction of our spacetime, called
$\ads$. The main ideas can also be read off from in the figures.  In
\sref{math}, we begin by introducing a certain set of coordinates on
anti-de~Sitter space, called $\ads_0$. They allow a very simple
description of light rays, and turn out to be very useful for the
construction of $\ads$. In these coordinates, anti-de~Sitter space is
represented as the interior of a timelike cylinder whose constant time
slices are Klein discs. The boundary of this cylinder, called $\scri_0$,
represents lightlike and spacelike infinity, and is also referred to as
the conformal boundary of $\ads_0$.  We investigate lightlike geodesics
and find that a null plane in $\ads_0$ can be defined by the family of
all light rays that emerge from a point on $\scri_0$ (\fref{ads}).

We also study the group structure of anti-de~Sitter space, which is
quite useful for the description of geodesics, isometries, and Killing
vectors. In particular we are interested in lightlike isometries, the
analogue to null rotations in Minkowski space, and put some effort in
understanding their action on $\scri_0$ (\fref{iso}). With this
preparation, we perform the very construction of the spacetime $\ads$ in
\sref{cut+glue}.  It contains two massless particles traveling in
opposite directions on lightlike geodesics. Since such geodesics
traverse the whole space, from one side of the cylinder to the other, in
a finite coordinate time, the particles appear from infinity at some
time, pass each other at a non-zero distance, and disappear again at a
later time. In this sense, all the interesting physics takes place
within a finite coordinate time interval.

The gravitational field of a point particle in three dimensional
gravity, with or without cosmological constant, can be constructed by
cutting out a wedge from a maximally symmetric spacetime and identifying
its faces \cite{2+1-part,MatschullWelling}. In the case of a massless
particle, we use a wedge that is degenerate to a null half plane
\cite{DeserSteif1}, which is uniquely determined by the lightlike
worldline of the particle. We cut anti-de~Sitter space along such a
surface, and then we identify the points on its lower face with those on
its upper face according to the action of a null rotation. The result is
that an observer who passes through this cut surface is effectively
mapped backwards in time. To perform this construction for two particles,
we can easily arrange the cut surfaces in such a way that they do not
overlap (\fref{cut}).

In \sref{gott}, we concentrate on the causal structure of the spacetime
$\ads$. A convenient way to do this is first to analyze the causal
structure of its conformal boundary $\scri$. Since the cut surfaces
intersect with the boundary of the cylinder, and since an observer
crossing one of them is effectively mapped backward in time, one easily
sees how a lightlike or timelike curve winding around the boundary may
close. It is not difficult to locate the region to which all closed
timelike curves on $\scri$ are confined (\fref{ctc}). We then discuss
how this information can be used in order to find the causal structure
of the interior of $\ads$. We find that there is a chronology horizon in
$\ads$, behind which every point lies on a closed timelike curve. The
horizon can be constructed in a remarkably simple way. We just have to
evolve null planes from appropriate points at $\scri$ (figures \ref{fix}
and \ref{chr}). This is very different from the situation in the flat
Gott universe, where the location of the CTC region are quite cumbersome
\cite{Cutler}.

Finally, in \sref{wormhole}, we change our point of view, and interpret
the region containing the closed timelike curves as a singularity. It
turns out to be a naked singularity inside a timelike wormhole, which
connects two otherwise separate regions of spacetime.  The particles
themselves are passing from one region to the other, through the
wormhole. The interior of the wormhole is separated from one exterior
region by a black hole event horizon, and from the second exterior
region by a white hole event horizon (\fref{csl}).  The event horizon
emerges from a cusp, which is located on the spacelike geodesic that
connects the particles in the moment when they fall into the black hole,
respectively when they fall out of the white hole (\fref{hor}). The
properties of the wormhole, such as its horizon length and angular
velocity, can be computed as functions of the energy and the minimal
distance, or impact parameter, of the particles. To complete the
discussion, we shall also consider an extremal (\fref{ext}), and a
static black hole (\fref{stc}).

\section{Anti-de Sitter Space}
\label{math}
To introduce the notation, let us give a brief description of three
dimensional anti-de~Sitter space, denoted by $\ads_0$. It can be covered
by a global, cylindrical coordinate chart $(t,\ch,\ph)$, with $\ch\ge0$
and $\ph\equiv\ph+2\pi$. It has a constant negative curvature, and the
metric with signature $(-,+,+)$ is
\begin{equation}
  \label{ads-metric-1}
  \dd s^2 = \dd\ch^2 + {\sinh^2}\ch \, \dd\ph^2 
                     - {\cosh^2}\ch \, \dd t^2 . 
\end{equation}
A somewhat different coordinate system, which is more appropriate for
the description of light rays, is obtained by replacing the hyperbolic
radial coordinate $0\le\ch<\infty$ by an angle $0\le\qh<\pi/2$. The
relation between $\ch$ and $\qh$ can be written in as
\begin{equation}
  \label{ch-qh}
  \tan\qh = \sinh\ch , \qquad
  \sin\qh = \tanh\ch , \qquad
  \cos\qh \, \cosh\ch = 1 ,
\end{equation}
which are all equivalent.  Inserting this into \eref{ads-metric-1} gives
\begin{equation}
  \label{ads-metric-2}
  \dd s^2 = \frac{\dd\qh^2 + {\sin^2}\qh \, \dd\ph^2 - \dd t^2}
                 {{\cos^2}\qh} 
\end{equation}
This tells us that anti-de~Sitter space is conformally isometric to the
direct product of a real time axis and a Euclidean half sphere of radius
one. With the sphere embedded in $\RR^3$, we can say that the conformal
factor is the inverse square of the $z$-coordinate. The conformally
transformed metric reads
\begin{equation}
  \label{conf-metric}
  \dd \tilde s^2 = \dd \qh^2 + {\sin^2}\qh \, \dd \ph^2 - \dd t^2.
\end{equation}
Including the equator as a boundary, we obtain a conformal
compactification of $\ads_0$. Its boundary is called $\scri_0$. It
represents spatial infinity as well as the origin and destination of
light rays.

\subsubsection*{Light rays}
As lightlike geodesics are not affected by conformal transformations, we
can say that the \emph{optical geometry} of anti-de~Sitter space is that
of a Euclidean half sphere \cite{Marek}. Every light ray travels along a
grand circle with a constant velocity of one. The fact that the boundary
$\scri_0$ is also a grand circle implies that every light ray on
$\ads_0$ intersects $\scri_0$ at two antipodal points, and the time that
it takes to travel from one side to the other is $\pi$. Moreover, all
light rays emerging from a point on $\scri_0$ at some time $t$ meet
again at the antipodal point at $t+\pi$. We call this a \emph{family} of
light rays.

\begin{figure}[t] 
  \begin{center}
    \epsfbox{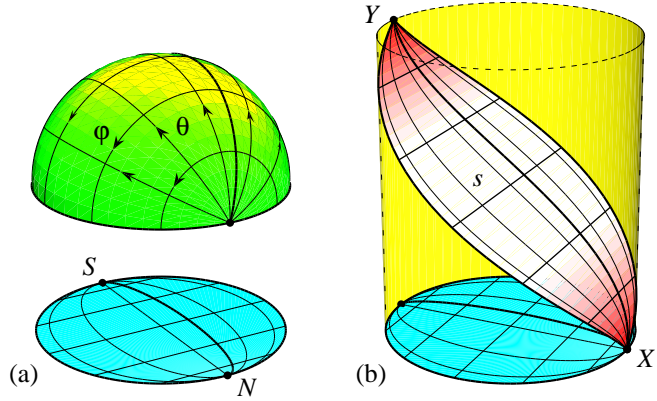}
    \caption{The spherical coordinates $(t,\qh,\ph)$ on the half
      sphere, the Klein disc, and the null plane $\srf$ spanned by the
      family of light rays connecting the antipodal points $X$ and $Y$
      on $\scri_0$. The bold line is the path of a light ray.}
    \label{ads}
  \end{center}
  \hrule
\end{figure}

With a slight modification, the spherical coordinate system can be
adapted to a special class of such families. Instead of representing the
slices of constant $t$ by the northern hemisphere, we take it to be the
eastern hemisphere. The latitude then runs from the north to the south
pole, such that $0<\qh<\pi$, and for the longitude we have $0<\ph<\pi$.
The metric \eref{ads-metric-2} is almost unchanged. Only the conformal
factor in front, which was formerly given by the inverse square of the
$z$-coordinate, now becomes the inverse square of the $y$-coordinate. In
terms of the rotated coordinates we have
\begin{equation}
  \label{ads-metric-3}
  \dd s^2 = \frac{\dd\qh^2 + {\sin^2}\qh \, \dd\ph^2 - \dd t^2 }
                 {{\sin^2}\qh\,{\sin^2}\ph} .
\end{equation}
The conformally transformed metric \eref{conf-metric} is unchanged. What
is quite useful to know is that the original hyperbolic coordinate $\ch$
is now related to the spherical coordinates by
\begin{equation}
  \label{ch-qh-ph}
  \cosh\ch = \frac1{\sin\qh \, \sin\ph}.
\end{equation}
The new coordinates are regular all over $\ads_0$, as can be seen in
\fref{ads}(a). The coordinate singularities at the poles $N$ and $S$ are
on the boundary. The lines of constant longitude $\ph$ are now the paths
of light rays connecting the poles, and the lines of constant latitude
$\qh$ at time $t$ are the wave fronts of the families of light rays that
started off from the north pole at $t-\qh$, and will arrive at the south
pole at $t-\qh+\pi$.

The disc below is obtained from the half sphere by orthogonal
projection. It is the so called \emph{Klein disc}, which is very closely
related to the \emph{Poincar\'e} disc \cite{BalaszVoros}. The latter has
already been used in previous work \cite{Matschull,SpinningAdS,ViBrill}.
It can be obtained by stereographic instead of orthogonal projection.
The special properties of the Klein disc, to which we shall stick within
this article, are not of particular importance. The only special feature
that is useful, but not necessary to know is that spatial geodesics are
represented as straight lines on the disc. This is the case, for
example, for the wave fronts belonging to a family of light rays, which
are the lines of constant latitude $\qh$. The main purpose we use the
Klein disc for is to draw three dimensional pictures of anti-de~Sitter
space.

We represent $\ads_0$ as a product of a Klein disc with a real line,
which becomes an infinitely long cylinder, whose boundary is $\scri_0$.
The time interval between $t=0$ and $t=\pi$ is shown in \fref{ads}(b).
The surface $\srf$ inside the cylinder is spanned by a family of light
rays.  They emerge from the north pole at $t=0$, denoted by $X$, and
arrive at the south pole at $t=\pi$, called $Y$. The surface is defined
by the simple coordinate equation $\qh=t$. The vertical lines on $\srf$
are the lines of constant $\ph$, representing the individual light rays.
The horizontal lines are those of constant $t$ and $\qh$, representing
the wave fronts at different times. The induced metric on the surface
can be expressed in terms of the coordinates $t$ and $\ph$. With $\qh=t$
in \eref{ads-metric-3}, we find that
\begin{equation}
  \label{srf-metric}
  \dd s = \dd\ph / \sin\ph. 
\end{equation}
It is of rank one because the surface is lightlike. The fact that it
only depends on $\ph$ implies that the light rays belonging to a family
are \emph{parallel}. Like in Minkowski space, a surface spanned by a
family of parallel light rays in anti-de~Sitter space is called a
\emph{null plane}. It can be regarded as the future or past light cone
attached to a point on the boundary $\scri_0$. We call $X$ the
\emph{origin} and $Y$ the \emph{destination} of the null plane $\srf$.
The part of anti-de~Sitter space above $\srf$ is the future of $X$, and
the part below $\srf$ is the past of $Y$. What is quite remarkable is
that this also defines a causal structure on $\scri_0$ itself.

The lines where the null plane intersects with the boundary of the
cylinder can be considered as two special light rays belonging to the
family, with $\ph=0$ and $\ph=\pi$. They travel counter clockwise,
respectively clockwise with a velocity of one along $\scri_0$. We can
also see this when we look at the conformally transformed metric on the
boundary. For both $\ph=0$ and $\ph=\pi$, the metric \eref{conf-metric}
becomes $\dd\tilde s^2=\dd\qh^2-\dd t^2$. This implies that there are
right and left moving light rays on $\scri_0$, with $|\dd\qh/\dd t|=1$.
This is a feature of anti-de~Sitter space that has no counterpart in
Minkowski space, where spacelike and lightlike infinity are not the
same, and lightlike infinity splits into two disconnected components,
one representing the origin and the other the destination of light rays.
Neither of them has an internal causal structure.

\subsubsection*{The group structure}
A very different way to think about anti-de~Sitter space is as a group
manifold. It is isometric to the matrix group $\grp$ of real $2\times2$
matrices with unit determinant. In terms of the spherical coordinates,
the isometry $\ads_0\to\grp$ is explicitly given by
\begin{equation}
  \label{ads-sl}
  \xx(t,\qh,\ph)  = 
      \frac{\cos t \, \one + \sin t \, \gam_0 
             + \cos\qh \, \gam_1 + \sin\qh \, \cos\ph \, \gam_2}
          {\sin\qh\,\sin\ph},
\end{equation}
where $\one$ is the unit matrix, and the gamma matrices form an
orthonormal basis of the associated Lie algebra $\alg$ of traceless
matrices,
\begin{equation}
  \label{gamma}
  \gam_0 = \pmatrix{ 0 & 1 \cr -1 & 0 }, \qquad
  \gam_1 = \pmatrix{ 0 & 1 \cr 1 & 0 }, \qquad
  \gam_2 = \pmatrix{ 1 & 0 \cr 0 & -1 }.
\end{equation}
One can check by straightforward calculation that the metric
\eref{ads-metric-3} is equal to the induced Cartan Killing metric
\begin{equation}
  \label{SL-metric}
  \dd s^2 = \ft12 \Trr{ \xx^{-1} \dd \xx \, \xx^{-1} \dd \xx },
\end{equation}
and therefore the map $\ads_0\to\grp$ is an isometry. However, the two
manifolds are only locally isometric. It is obvious from \eref{ads-sl}
that the coordinate $t$ becomes periodic on $\grp$. To obtain $\ads_0$
from $\grp$, one has to unwind its fundamental loop. Hence,
anti-de~Sitter space is the covering manifold of the group $\grp$. In
practice, we do not have to care about this very much, because all our
constructions will take place within a time interval of $\pi$, and
therefore the period of $2\pi$ in $t$ cannot be seen.

The probably most useful feature of the group structure is that it
allows a very simple description of isometries and Killing vectors. A
generic time and space orientation preserving isometry of $\grp$ can be
written as
\begin{equation}
  \label{sl-iso}
  \iso: \quad \xx \mapsto \uu^{-1} \xx \, \vv,
\end{equation}
where $\uu$ and $\vv$ are two arbitrary group elements. If
$\uu=\expo{\mm}$ and $\vv=\expo{\nn}$ are exponentials of some elements
$\mm,\nn\in\alg$, then the action of $\iso$ can be written as the flow
of a Killing vector field $\kll$. This can be defined by
\begin{equation}
  \label{kll-def}
  \kll (\xx) = \xx \, \nn - \mm \, \xx, 
\end{equation}
where $\xx$ is considered as an $\grp$ valued function on $\ads_0$. Note
that this determines $\kll$ uniquely, and implies that for the flow of
$\kll$, denoted by $\expo{\tau\kll}$, $\tau\in\RR$, we have
\begin{equation}
  \label{kll-exp}
  \expo{\tau\kll} (\xx) = \expo{-\tau\mm} \, \xx \, \expo{\tau\nn}.
\end{equation}
For $\tau=1$, this is equal to \eref{sl-iso}. Given a second isometry
$\riso:\xx\mapsto\gx^{-1}\xx\hx$, $\gx,\hx\in\grp$, we can compute its
action on the vector field $\kll$. For the push forward $\riso^*(\kll)$,
we must have 
\begin{equation}
  \expo{\tau\riso^*(\kll)} 
     = \riso \circ \expo{\tau\kll} \circ \riso^{-1},
\end{equation}
which implies that 
\begin{equation}
  \label{kll-push}
  \riso^*(\kll)(\xx)  = \xx \, ( \hx^{-1} \nn \, \hx ) 
                       - ( \gx^{-1} \mm \, \gx ) \,  \xx.
\end{equation}
Hence, the two vectors $\mm,\nn\in\alg$ transform independently under
the adjoint representation of $\gx,\hx\in\grp$. If we interpret them as
vectors in three dimensional Minkowski space, which is isometric to
$\alg$, then the transformations 
\begin{equation}
  \label{lorentz}
  \riso^* : \quad
       \mm \mapsto \gx^{-1} \mm \, \gx, \quad
       \nn \mapsto \hx^{-1} \nn \, \hx,
\end{equation}
can be considered as proper Lorentz transformations in $\alg$. They
preserve the invariant lengths $\ft12{\Tr(\mm^2)}$ and
$\ft12{\Tr(\nn^2)}$ of $\mm$ and $\nn$. Given two Killing vectors
$\kll_1$ and $\kll_2$, we can also compute their scalar product using
the formula \eref{SL-metric},
\begin{eqnarray}
  \label{kll-prod}
  \kll_1\cdot\kll_2 
   &=& \ft12\Trr{ \xx^{-1}\,\kll_1(\xx) \, \xx^{-1}\,\kll_2(\xx)} \nwl 
   &=& \ft12\Trr{ \nn_1 \, \nn_2 + \mm_1 \, \mm_2 
                - \xx^{-1} \mm_1 \, \xx \, \nn_2 
                - \xx^{-1} \mm_2 \, \xx \, \nn_1 }.
\end{eqnarray}
This is a scalar function on $\ads_0$. Its limit on the boundary
$\scri_0$ will in general diverge, because of the zero denominator in
\eref{ads-sl}. However, under a certain condition, namely
\begin{equation}
  \label{kll-ortho}
  \alpha \, \mm_1 = \beta \, \mm_2 , \qquad
  \alpha \, \nn_1 = - \beta \, \nn_2, \qquad \alpha,\beta \in\RR,
\end{equation}
the scalar product is constant, equal to the sum of the scalar products
of the Minkowski vectors $\mm$ and $\nn$, and it remains finite at the
boundary. We then call the two Killing vectors \emph{asymptotically
  orthogonal}, or orthogonal on $\scri_0$. In fact, if we first derive
the limits of $\kll_1$ and $\kll_2$ on $\scri_0$, and then take their
scalar product with respect to the conformally transformed metric
\eref{conf-metric}, then they turn out to be orthogonal.

\subsubsection*{Geodesics}
Another useful property of isometries, or Killing vectors on the group
manifold is that their fixed points, if there are any, always form a
geodesic. We call this the \emph{axis} of the isometry. Vice versa,
every geodesic is the axis of some isometry. This is again similar to
Minkowski space, where every straight line is the axis of a one
parameter family of Poincar\'e transformations. To see that this is also
the case in anti-de~Sitter space, consider the fixed point equation for
the isometry $\iso$ defined above, respectively its generating Killing
vector $\kll$ in the form
\begin{equation}
  \label{fp}
  \uu \, \xx = \xx \, \vv \equivalent
  \mm \, \xx = \xx \, \nn.
\end{equation}
It can be solved if and only if $\uu$ and $\vv$ lie in the same
conjugacy class of $\grp$, or equivalently if the vectors $\mm$ and
$\nn$ in $\alg$ are related by a proper Lorentz transformation. Let us
assume this, and that it is not one of the trivial classes
$\uu=\vv=\pm\one$, or $\mm=\nn=0$. Otherwise there are either no fixed
points at all, or $\iso=\id$ and $\xi=0$. To show that the solutions to
\eref{fp} form a geodesic, we pick a particular solution $\yy$ and
replace the variable $\xx$ by $\zz=\xx\yy^{-1}$. The relation between
$\xx$ and $\zz$ is again an isometry. It is therefore sufficient to show
that the solutions for $\zz$ lie on a geodesic. The equation for $\zz$
can then be written as
\begin{equation}
  \uu \, \zz  = \zz \, \uu
  \equivalent
  \mm \, \zz  = \zz \, \mm ,
\end{equation}
where we have used that $\uu\yy=\yy\vv$, or equivalently
$\mm\yy=\yy\nn$. Hence, $\zz$ is required to commute with $\mm$, or its
exponential $\uu$. In a Lie group of rank one, this is the case for all
$\zz$ belonging to the unique one-dimensional subgroup that contains
$\uu$, which is that generated by the exponential of the element $\mm$
of the algebra. On the other hand, a one dimensional subgroup of a Lie
group is a geodesic passing through the unit element. This proves that
the fixed points form a geodesic.

To see that every geodesics is the axis of some isometry, take some
geodesic and choose two distinct points $\yy$ and $\zz$ thereon, such
that $\uu=\zz\yy^{-1}$ and $\vv=\yy^{-1}\zz$ are different from
$\pm\one$. The fixed point equation \eref{fp} is then solved by $\yy$
and $\zz$, which means that the axis of this isometry is the unique
geodesic passing through $\yy$ and $\zz$. There is a one parameter
family of such isometries, as can be seen by varying $\zz$ and $\yy$
along the geodesic. A priory, this provides two parameters, but one can
easily convince oneself that a variation of $\yy$ can always be
compensated by a variation of $\zz$ and vice versa, so that actually
there is only one parameter. As in Minkowski space, this can be regarded
as the \emph{angle of rotation}.

\subsubsection*{Null isometries}
Whether the axis of an isometry is a timelike, lightlike, or spacelike
geodesic depends on the conjugacy class to which $\uu$ and $\vv$ belong.
The lightlike ones are of particular interest for us. As an example,
consider the following two positive lightlike group elements,
\begin{equation}
  \label{p-par}
  \uu = \one + \expo{-\dis} \, \tan\erg \, (\gam_0 - \gam_2 ), \qquad
  \vv = \one + \expo{\dis}  \, \tan\erg \, (\gam_0 + \gam_2 ),
\end{equation}
where $\dis$ and $0<\erg<\pi/2$ are two real parameters. Both $\uu$ and
$\vv$ lie on the future light cone of the unit element $\one\in\grp$,
belonging to the same conjugacy class. The reason for choosing the
parameters in this particular way will become clear later on. We expect
that the axis of the corresponding \emph{null isometry} is a lightlike
geodesic. Using the representation of $\xx$ in terms of the spherical
coordinates \eref{ads-sl}, the fixed point equation becomes
\begin{equation}
  \label{p-fp}
  \uu \, \xx = \xx \, \vv \equivalent
  \cos\qh = \cos t , \quad
  \sin\qh \, \cos\ph = \sin t \, \tanh\dis.
\end{equation}
The first equation can always be solved for $\qh$, which ranges from $0$
to $\pi$. It tells us that the geodesic is oscillating with a period of
$2\pi$ between the north and the south pole. As these points do not
belong to $\ads_0$, the curve actually splits into a series of
geodesics, each within a time interval of $\pi$. For $0<t<\pi$, the
coordinate equations can then be simplified to
\begin{equation}
  \label{p-fp-0}
  \uu \, \xx = \xx \, \vv \equivalent
  \qh = t , \quad
  \cos\ph = \tanh\dis.
\end{equation}
Obviously, this is one of the light rays belonging to the family
considered above.  The first equation defines the null plane $\srf$ in
\fref{ads}(b), and the second equation picks out a particular light ray.
The parameter $\dis$ increases from $-\infty$ in the front of the
picture to $+\infty$ in the back. In the limits $\dis\to\pm\infty$, we
also recover the two light rays traveling along $\scri_0$. What is
useful to know is that $\dis$ measures the physical distance between the
individual light rays. To see this, we have to express the metric
\eref{srf-metric} on the null plane in terms of $\dis$ instead of $\ph$,
\begin{equation}
  \label{srf-metric-dis}
  \dd s = \dd\ph/{\sin\ph}
        = - \dd\cos\ph/{\sin^2}\ph
        = - {\cosh^2}\dis \, \dd\tanh\dis 
        = - \dd\dis. 
\end{equation}
The minus sign appears because $\ph$ decreases with increasing $\dis$.
Coming back to the isometry specified by $\dis$ and $\erg$, we can say
that the parameter $\dis$ determines the location of the axis of $\iso$,
and we expect $\erg$ to specify the angle of rotation.

In Minkowski space, the action of a lightlike rotation on the unique
null plane that contains the axis is such that each light ray in that
plane is a fixed line. The points on the null plane are shifted along
the light rays by an amount that is proportional to the angle of
rotation and the distance from the axis. To see that this is also the
case in anti-de~Sitter space, let us make the following ansatz.  On the
surface $\srf$, we introduce the coordinates $t$ and $\ph$, such that a
point with spherical coordinates $(t,t,\ph)$ is simply denoted by
$(t,\ph)$. It is then represented by the group element
\begin{equation}
  \label{surf-sl}
  \xx(t,\ph) = 
       \frac{ \cot t \, (\one + \gam_1) 
            + \gam_0 + \cos\ph \, \gam_2  }
             {\sin\ph} ,
\end{equation}
which is obtained from \eref{ads-sl} with $\qh=t$. Now, assume that
$\iso$ maps a point $\xx_-=\xx(t_-,\ph)$ onto $\xx_+=\xx(t_+,\ph)$ on
the same light ray. This yields the following relation between the
time coordinates,
\begin{equation}
  \label{surf-iso}
  \xx_- \, \vv = \uu \, \xx_+  \equivalent
  \cot t_+ - \cot t_- = 2 \, \tan\erg \, 
                  ( \cosh\dis \, \cos\ph - \sinh\dis) .
\end{equation}
For $0<t<\pi$, this provides a one-to-one relation between $t_-$ and
$t_+$. From this we conclude that the ansatz was correct. The null plane
$\srf$ is indeed a fixed surface of $\iso$, and the individual light
rays belonging to the family are fixed lines. The points are shifted
along the light rays by an amount that increases with the angle of
rotation and the distance from the axis. Note that on the axis we have
$\cos\ph=\tanh\dis$, and therefore $t_+=t_-$. Apart from the fact that
the amount of shift is not proportional to the parameters, the situation
is the same as in Minkowski space.

We can also ask the reverse question. Given a light ray, what is the
most general isometry, or Killing vector, that has this light ray as a
fixed line? Or, more generally, what is the most general isometry that
has a fixed light ray? This will be useful to know at several points in
the derivation of the physical properties of our wormhole spacetime. The
first observation we can make is that whenever there is a fixed light
ray of some isometry $\iso$, then all members of its family are fixed
lines of $\iso$ as well. The argument is the same as above. The null
plane containing the given light ray is a fixed surface of $\iso$,
because there is only one such null plane, and the physical distances
between the members of the family are preserved. So, the actual question
is, which isometries have a family of fixed light rays?

Let us give an explicit answer for the special family of light rays
spanning the null plane $\srf$ defined above, and then generalize the
result. As an ansatz, consider the action of a Killing vector $\kll$
defined by \eref{kll-def}. To see whether $\kll$ is tangent to the given
family of light rays, we have to evaluate $\kll(\xx)$ on $\srf$, express
this as a function of the coordinates $(t,\ph)$ on $\srf$, as defined by
\eref{surf-sl}, and check whether the result is proportional to
$\dd\xx/\dd t$. It turns out that this is the case if and only if the
Minkowski vectors $\mm$ and $\nn$ defining the Killing vector $\kll$ are
of the form
\begin{equation}
  \label{kll-tng-par}
  \mm = a \, (\gam_0 - \gam_2) + c \, \gam_1, \qquad
  \nn = b \, (\gam_0 + \gam_2) - c \, \gam_1,
\end{equation}
with real numbers $a,b,c$. Explicitly, the action of the Killing vector
$\kll$ on the light rays in $\srf$ is then given by 
\begin{equation}
  \label{kll-tng-act-t}
  \dd \cot t = (a+b) \, \cos\ph + (a-b) - 2 \, c \, \cot t.
\end{equation}
With $a=\expo{-\dis}\tan\erg$, $b=\expo{\dis}\tan\erg$, and $c=0$, we
can easily integrate this and recover the action \eref{surf-iso} of
$\iso=\expo{\kll}$. We then also have $\uu=\expo{\mm}$ and
$\vv=\expo{\nn}$, with $\uu$ and $\vv$ given in \eref{p-par}.

To generalize this result, we have to find the properties which are
common to all Minkowski vectors \eref{kll-tng-par}, and which are
invariant under isometries of anti-de~Sitter space. We know that under a
general isometry $\riso:\xx\mapsto\gx^{-1}\xx\hx$, the vectors $\mm$ and
$\nn$ transform under proper Lorentz rotations \eref{lorentz}.  The only
invariants under such transformations are the length of $\mm$ and $\nn$,
and in the case of lightlike and timelike vectors we have to distinguish
between positive and negative ones. Now, it turns out that the vectors
given above are always lightlike or spacelike and of the same length,
\begin{equation}
  \label{kll-tng-con}
  \Trr{\mm^2} = \Trr{\nn^2} \ge 0 .
\end{equation}
Moreover, if they are lightlike, then we can have any combination of
positive and negative lightlike vectors. Hence, we expect that
\eref{kll-tng-con} is a necessary and sufficient condition for a Killing
vector to have a family of fixed light rays. Indeed, given a Killing
vector $\kll$ with $\mm$ and $\nn$ fulfilling \eref{kll-tng-con}, then
we can always find an isometry $\riso:\xx\mapsto\gx^{-1}\xx\hx$ such
that $\gx^{-1}\mm\gx$ and $\hx^{-1}\nn\hx$ are of the form
\eref{kll-tng-par}, for some parameters $a,b,c$. The Killing vector
$\riso^*(\kll)$ is then tangent to the light rays in $\srf$, which
implies that $\kll$ has a family of fixed light rays, namely those
spanning the null plane $\riso^{-1}(\srf)$.

\subsubsection*{Isometries acting on $\scri_0$}
\begin{figure}[t]
  \begin{center}
    \epsfbox{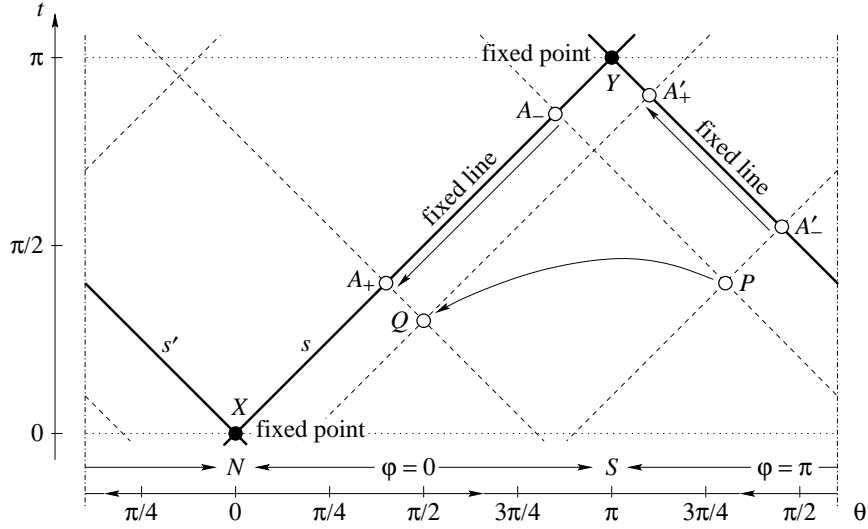}
    \caption{The action of a null isometry on $\scri_0$. The fixed lines
      $\srf$ and $\srf'$ are the intersections of the null plane in
      \fref{ads} with the boundary of the cylinder. The fixed points $X$
      and $Y$ are the end points of the axis. The image $Q=\iso(P)$ of a
      point $P$ on $\scri_0$ can be constructed when the action of
      $\iso$ on the fixed light rays is known.}
    \label{iso}
  \end{center}
  \hrule
\end{figure}

Finally, let us also give a brief description of the action of a null
isometry on the conformal boundary $\scri_0$ of $\ads_0$. If we cut the
boundary of the cylinder in \fref{ads} along a vertical line and lay it
down on a plane, we get an infinitely long strip. Again for $0<t<\pi$,
this is shown in \fref{iso}.  To recover the cylinder surface, the left
and right margins must be identified, so that everything that leaves the
strip to the right reappears on the left and vice versa. A disadvantage
of our spherical coordinate system is that it does not provide proper
coordinates on $\scri_0$, because the poles are located there. However,
for most of the constructions to be made on $\scri_0$, it is sufficient
to use the time coordinate $t$.

The bold lines denoted by $\srf$ and $\srf'$ in \fref{iso} are the
intersections of the null plane in \fref{ads}(b) with the boundary of
the cylinder. They represent a left and a right moving light ray on
$\scri_0$. We already know that these are fixed lines of $\iso$. If
$A_-$ and $A_+$ are two points on $\srf$ such that $A_+=\iso(A_-)$, then
their time coordinates satisfy
\begin{equation}
  \label{scri-iso-1}
  \cot t_+ - \cot t_- = 2 \, \expo{-\dis} \, \tan\erg. 
\end{equation}
This follows from \eref{surf-iso} with $\ph=0$. Similarly, if $A'_-$ and
$A'_+$ are two points on $\srf'$, such that $A'_+=\iso(A'_-)$, then the
relation between their time coordinates is 
\begin{equation}
  \label{scri-iso-2}
  \cot t'_+ - \cot t'_- = - 2 \, \expo{\dis} \, \tan\erg, 
\end{equation}
which is obtained from \eref{surf-iso} with $\ph=\pi$. It is useful to
keep in mind what the directions of these transformations are. For
positive $\erg$, points on $\srf$ are shifted \emph{downwards}, and
those on $\srf'$ are shifted \emph{upwards}.

The intersections $X$ and $Y$ of the two light rays are the fixed points
of $\iso$. This is where the axis of $\iso$ intersects with $\scri_0$.
Quite generally, a fixed point of an isometry on $\scri_0$ is always at
the intersection of a left and a right moving fixed light ray, and vice
versa the two light rays passing through a fixed point are fixed lines.
If we extend the light rays $\srf$ and $\srf'$, we find that $\iso$ has
more fixed points. There are two intersections within each time period
of $2\pi$. However, as already mentioned, all the interesting physics
will take place within the time interval $0<t<\pi$, and therefore it is
sufficient to consider this region only.

Knowing the action of $\iso$ on the fixed light rays $\srf$ and $\srf'$,
we can easily construct the image $Q=\iso(P)$ of any point $P$ on
$\scri_0$.  The fixed light rays divide $\scri_0$ into a series of
\emph{diamonds}.  The continuity of $\iso$ implies that $P$ and $Q$ lie
in the same diamond.  Now, consider the two light rays passing through
$P$ and denote their intersection with the fixed lines by $A_-$ and
$A'_-$, as indicated in \fref{iso}. The images of these light rays are
those that pass through $A_+=\iso(A_-)$ and $A'_+=\iso(A'_-)$,
determined by \eref{scri-iso-1} and \eref{scri-iso-2}. It follows that
the image of $P$ is at the only intersection $Q$ of the light rays
through $A_+$ and $A'_+$ within the same diamond.

\section{The Spacetime}
\label{cut+glue}
We are now ready to construct our spacetime manifold $\ads$. What we are
looking for is a solution to Einstein's equations with a negative
cosmological constant and two massless, pointlike particles as matter
sources. In three dimensions, such a spacetime has a constant negative
curvature everywhere, except for two conical singularities located on
the world lines of the particles. It can be constructed from
anti-de~Sitter space $\ads_0$ by \emph{cutting} and \emph{gluing}
\cite{2+1-part,DeserJackiw}.  For each particle, one has to choose a
world line and a so called \emph{wedge}. The wedge is a region of
$\ads_0$ which is bounded by two \emph{faces}, that is, two surfaces
extending from the world line to infinity, such that one of them is
mapped onto the other by an isometry, whose axis is the world line.

Taking away the interior of the wedges and identifying the points on the
faces according to the corresponding isometries, one obtains a spacetime
$\ads$ which is locally isometric to $\ads_0$, but with conical
singularities located on the world lines. This method has already been
used to describe the collapse of two colliding particles into a static
black hole \cite{Matschull}. Here, we want to generalize it to the case
of two non-colliding particles approaching each other and collapsing
into a rotating black hole. It turns out that in this case the black
hole is actually a timelike wormhole with a naked singularity therein.
The particles pass through the wormhole into a second exterior region.
To begin with, let us give a definition of the spacetime $\ads$.

\subsubsection*{Massless particles}
There are two physically relevant parameters describing the relative
motion of two massless point particles passing each other. Namely, their
distance at the moment of closest approach and their centre of mass
energy. The distance is the length of the unique spacelike geodesic that
is orthogonal to both lightlike world lines. For two identical
particles, the centre of mass frame can be defined as follows. It is
that coordinate system $(t,\qh,\ph)$ in which the particles are
interchanged by reflection at the time axis, or rotation of the cylinder
by $180$ degrees,
\begin{equation}
  \label{rot-def}
  \rot: \quad (t,\qh,\ph) \mapsto (t,\pi-\qh,\pi-\ph), 
        \quad \xx \mapsto \gam_0\!^{-1} \, \xx \, \gam_0.
\end{equation}
Note that the reflection is an involution, $\rot\circ\rot=\id$, but not
a parity transformation. Within the centre of mass frame, we still have
the freedom to perform time shifts and spatial rotations. Using this, we
can achieve that the spacelike geodesic joining the particles at the
moment of closest approach is the coordinate line $t=\qh=\pi/2$. This
the horizontal line in the centre of the null plane in \fref{ads}(b).
One of the world lines is then a member of the family spanning this
surface, and the second one is obtained by reflection.  Both can be
specified as the axes of appropriate isometries,
\begin{equation}
  \label{p-iso}
  \iso_1: \quad \xx \mapsto \uu_1\inv \xx \, \vv_1 , \qquad
  \iso_2: \quad \xx \mapsto \uu_2\inv \xx \, \vv_2 . 
\end{equation}
The parameters for the first particle are those already introduced in
\eref{p-par},
\begin{equation}
  \label{p1-par}
  \uu_1 = \one + \expo{-\dis} \tan\erg \, (\gam_0 - \gam_2 ), \qquad
  \vv_1 = \one + \expo{\dis}  \tan\erg \, (\gam_0 + \gam_2 ).
\end{equation}
The second isometry is determined by $\rot\circ\iso_2=\iso_1\circ\rot$,
which implies that
\begin{equation}
  \label{p2-par}
  \uu_2 = \one + \expo{-\dis} \tan\erg \, (\gam_0 + \gam_2 ), \qquad
  \vv_2 = \one + \expo{\dis}  \tan\erg \, (\gam_0 - \gam_2 ).
\end{equation}
The coordinate equations for the world lines are
\begin{equation}
  \label{p-lines}
  \prt_1:\quad \qh = t,\quad \cos\ph=\tanh\dis, \qquad
  \prt_2:\quad \qh = \pi-t, \quad \cos\ph=-\tanh\dis.
\end{equation}
From the results of the previous section, it follows that the physical
distance between the particles at the moment of closest approach is
$2\dis$. The second parameter $\erg$ measures the centre of mass energy
of the particles. It specifies the angle of rotation of $\iso_1$ and
$\iso_2$, and therefore the \emph{size} of the wedges to be cut out. It
is not so important what the precise definition of the centre of mass
energy in a curved spacetime is, and what the relations are between
$\erg$, the momenta of the particles and the properties of the conical
singularities \cite{Matschull,2+1-part,MatschullWelling}. It is
sufficient to know that, at least qualitatively, $\erg$ measures the
amount of energy contained in the combined system.

\subsubsection*{The cut surfaces}
\begin{figure}[t]
  \begin{center}
    \epsfbox{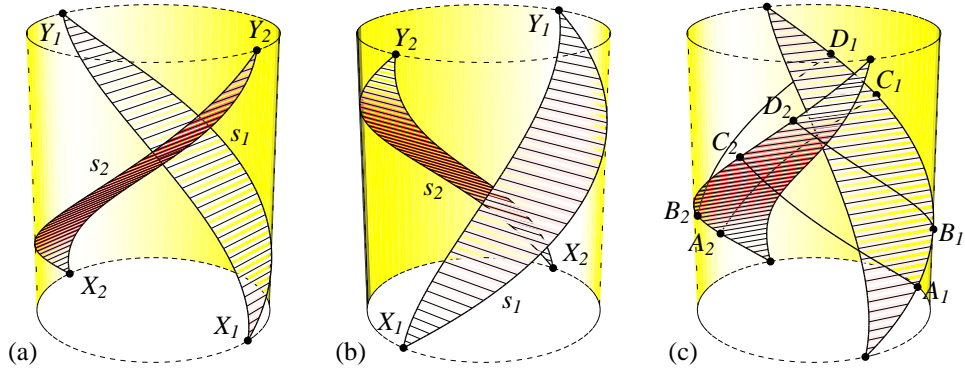}
    \caption{The spacetime $\ads$ is constructed from $\ads_0$ by cutting 
      and gluing along the null half planes $\srf_1$ and $\srf_2$ (a,b).
      They extend from the world lines $\prt_1$ and $\prt_2$ to the
      boundary. The solid lines (c) on the boundary are the two closed
      light rays constructed in \fref{ctc}. Here and in all following
      figures, the lines drawn on surfaces are lines of constant time
      coordinate $t$.}
    \label{cut}
  \end{center}
  \hrule
\end{figure}

For massless particles moving on lightlike world lines, the wedges can
be chosen in a special way. Let us from now on assume that $\dis>0$, and
consider the following \emph{null half planes},
\begin{equation}
  \label{p-planes}
  \srf_1: \quad \qh = t,\quad \cos\ph\ge\tanh\dis, \qquad
  \srf_2: \quad \qh = \pi-t, \quad \cos\ph\le-\tanh\dis.
\end{equation}
These are two non-overlapping lightlike surfaces extending from the
world lines to $\scri_0$, as shown in \fref{cut}. We already know that
the half plane $\srf_1$ is a fixed surface of $\iso_1$, and due to the
symmetry under reflection $\srf_2$ is a fixed surface of $\iso_2$. On
both surfaces, we introduce coordinates $t$ and $\ph$, such that a point
$(t,t,\ph)\in\srf_1$ is denoted by $(t,\ph)$, and the rotated point
$(t,\pi-t,\pi-\ph)\in\srf_2$ is also denoted by $(t,\ph)$. When written
in these coordinates, the action of $\iso_2$ on $\srf_2$ is formally the
same as that of $\iso_1$ on $\srf_1$, namely
\begin{equation}
  \label{srf-iso}
  \iso_{1,2}: \quad (t_-,\ph) \mapsto (t_+,\ph), \qquad
  \cot t_+ - \cot t_- = 2 \, \tan\erg \, 
                  ( \cosh\dis \, \cos\ph - \sinh\dis ). 
\end{equation}
To construct the spacetime $\ads$, we consider the half planes $\srf_1$
and $\srf_2$ as two \emph{degenerate} wedges. There is then no interior
of the wedges to be removed. Instead, the points on the upper and the
lower faces of the null planes are considered as distinct. The
identification of the faces is such that a point $(t_-,\ph)$ on the
\emph{lower} face corresponds to the point $(t_+,\ph)$ on the
\emph{upper} face of the same null half plane. Note that for a lightlike
surface there is a well defined distinction between the upper and lower
face, the former being the one that points towards the future, and the
latter being that pointing towards the past.

The resulting spacetime $\ads$ is still covered by a single spherical
coordinate chart $(t,\qh,\ph)$, but now in a non-trivial way. The
coordinates are discontinuous along the cut surfaces. They can be
considered as self-overlap regions, with the transition functions given
by \eref{srf-iso}. As a result, $\ads$ contains two conical
singularities located on lightlike world lines of massless particles.
Their centre of mass energy is $\erg$, and they pass each other at a
distance of $2\dis$. Perhaps we should also mention that the degenerate
wedges are very similar to the gravitational shock waves carried by
massless particles in higher dimensions \cite{Penrose,AichSexl}. The
gravitational field of such a particle can also be constructed by
cutting and gluing.

The cut surface is thereby also the unique null hyperplane that contains
the world line. However, the map that provides the identification is not
an isometry of the underlying symmetric space, which is either Minkowski
or anti-de~Sitter space. As a consequence, additional curvature in form
of a gravitational shock wave is introduced on the cut surface. This is
not the case in three dimensions. Here, it is possible to choose the
coordinates such that the cut surface becomes a \emph{half} plane,
attached to one side of the world line only. This is quite essential,
because otherwise the two cut surfaces would overlap, and the cutting
and gluing procedure would not work for two particles simultaneously. It
is also a nice way to see that there is no interaction between the
particles by local forces.

\section{The Gott Universe}
\label{gott}
Regarding the construction of the spacetime manifold, we are now already
finished. This is probably a bit surprising. It is not at all obvious
that there are any of the typical features of a wormhole, such as a
singularity, horizons, or a kind of tunnel connecting otherwise separate
regions of spacetime. Moreover, it seems that there cannot be any
singularity at all, except for the harmless ones on the world lines,
because the spacetime is constantly curved. In fact, we have to
introduce the wormhole singularity in a somewhat artificial way, which
is however quite standard in the construction of three dimensional black
holes.  Before doing so, let us have a closer look at our spacetime
manifold $\ads$ as it is, from a slightly different point of view.

What we have constructed is the anti-de~Sitter analogue of the
\emph{Gott universe} \cite{Gott,DeserSteif2}. The Gott universe is a
spacetime with two point particles and vanishing cosmological constant.
It can be constructed from Minkowski space by essentially the same
cutting and gluing procedure.  Provided that the centre of mass energy
of the particles exceeds a certain threshold, it contains closed
timelike curves. They fill a region that extends from spatial infinity
to some neighbourhood of the particles at the moment of closest
approach. That there is a certain danger for this to happen in our
spacetime as well can be inferred from the relation \eref{srf-iso}. The
essential point is that the points on the lower faces with time
coordinate $t_-$ are mapped onto points on the upper faces with a
\emph{smaller} time coordinate $t_+$.

As a consequence, a future pointing timelike curve, approaching one of
the cut surfaces from below, continues above the cut surface at an
earlier time. We can say that when passing over the cut surfaces, we
\emph{gain} time. Doing so several times, it might be possible to form a
closed timelike curve. This is indeed what is going to happen, provided
that the energy of the particles is large enough, and, in contrast to
the Gott universe where this is already sufficient, the distance between
the particles has to be small. In other words, a sufficiently large
amount of energy has to be located in a small volume. This is a typical
condition for a black hole to be created. The purpose of this section is
to find the exact condition to be imposed on the energy and the
distance, and to determine the subset of $\ads$ which is filled by the
closed timelike curves.
\begin{figure}[t]
  \begin{center}
    \epsfbox{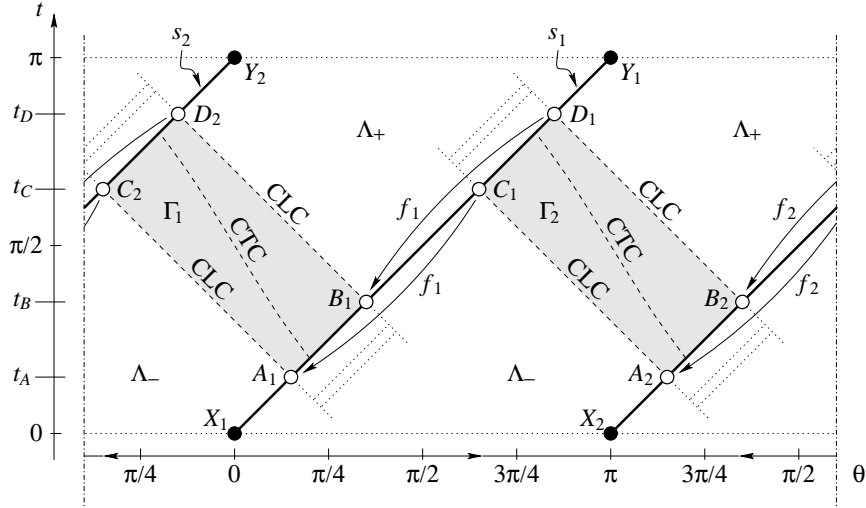}
    \caption{The boundary $\scri$ of $\ads$ is obtained from the
      boundary $\scri_0$ of $\ads_0$ by cutting and gluing along the
      right moving light rays $\srf_1$ and $\srf_2$. The left moving
      light rays $A_1C_2A_2C_1$ and $B_1D_2B_2D_1$ are closed lightlike
      curves on $\scri$. The shaded region $\sng=\sng_1\cup\sng_2$ is
      filled with closed timelike curves.}
    \label{ctc}
  \end{center}
  \hrule
\end{figure}

\subsubsection*{The causal structure of $\scri$}
A convenient way to find the condition for closed timelike curves to
arise is to look at the conformal boundary $\scri$ of $\ads$.  This can
be constructed from the boundary $\scri_0$ of $\ads_0$ by the following
cutting and gluing procedure. \Fref{ctc} represents the boundary of the
cylinder in \fref{cut}. The bold lines are the intersections of the cut
surfaces $\srf_1$ and $\srf_2$ with the boundary. They extend from the
points $X_{1,2}$ at $t=0$, where the particles emerge, to the points
$Y_{1,2}$ at $t=\pi$, where they disappear again. Along these lines the
boundary is cut. The identification is such that a point with time
coordinate $t_-$ below one of the cuts is identified with a point with
time coordinate $t_+$ above the cut. According to \eref{scri-iso-1}, the
relation between $t_-$ and $t_+$ is
\begin{equation}
  \label{clc-jump}
  \cot t_+ - \cot t_- = 2 \, \expo{-\dis} \, \tan\erg . 
\end{equation}
When the cut is traversed from below, this is always a shift backward in
time, $t_+<t_-$. The crucial question is whether a left moving light ray
on $\scri$ can gain enough time by passing over the cuts to form a loop.
The time that it takes for such a light ray to travel from one cut to
the other is always $\pi/2$. This must be equal to the time gained when
passing over the cut, $t_--t_+=\pi/2$. Inserting this into
\eref{clc-jump}, we find that
\begin{equation}
  \label{clc-con}
  \sin( 2 \, t_+ ) = - \sin( 2 \, t_- ) = \expo{\dis} \cot\erg.
\end{equation}
There is no solution if the right hand side is bigger than one. The time
gained is then always smaller then $\pi/2$, and therefore it is not
possible for a light ray to form a closed loop. In particular, this is
the case for small energies or large distances, which is quite
reasonable. In fact, in both limits $\erg\to0$ and $\dis\to\infty$, the
particles disappear and $\ads$ becomes equal to empty anti-de~Sitter
space $\ads_0$.

For closed light rays on $\scri$ to exist, the energy must be bigger
than $\pi/4$, so that $\tan\erg>1$. This is also the threshold for a
static black hole to be formed by colliding particles \cite{Matschull}.
Here we get the additional condition that the spatial separation of the
particles must be sufficiently small, $\expo{\dis}<\tan\erg$. Then we
have the situation shown in \fref{ctc}. There are two closed light rays
$A_1C_2A_2C_1$ and $B_1D_2B_2D_1$, where 
\begin{equation}
  \label{clc-points}
  A_1=\iso_1(C_1), \quad
  B_1=\iso_1(D_1), \qquad
  A_2=\iso_2(C_2), \quad
  B_2=\iso_2(D_2),
\end{equation}
and for the time coordinates we have
\begin{equation}
  \label{clc-dif}
  t_C=t_A+\pi/2, \qquad t_D=t_B+\pi/2.
\end{equation}
To determine the location of these points explicitly, it is useful to
replace the parameters $\dis$ and $\erg$ by two positive real numbers
$0<\mu<\nu$, such that
\begin{equation}
  \label{mu-nu}
  \cosh(\mu/2) = \expo{-\dis} \, \tan\erg , \qquad
  \cosh(\nu/2) = \expo{\dis} \, \tan\erg .
\end{equation}
Note that the right hand sides of both equations are bigger or equal to
one if $\dis\ge0$ and $\expo{\dis}\le\tan\erg$. We can then solve the
equations \eref{clc-con} and \eref{clc-dif}, and find that 
\begin{equation}
  \label{clc-times}
  \cot t_A = \expo{\mu/2}, \quad
  \cot t_B = \expo{-\mu/2}, \quad
  \cot t_C = -\expo{-\mu/2}, \quad
  \cot t_D = -\expo{\mu/2}.
\end{equation}
An alternative and rather elegant way to characterize the light rays
that form the closed loops on $\scri$ is the following. When considered
as light rays on $\scri_0$, that is, before the cutting and gluing
procedure is carried out, they are fixed lines of certain combinations
of the isometries $\iso_1$, $\iso_2$ and the reflection $\rot$.  Let us
define
\begin{equation}
  \label{riso}
  \riso_1 = \iso_1 \circ \rot = \rot \circ \iso_2, \qquad
  \riso_2 = \iso_2 \circ \rot = \rot \circ \iso_1.
\end{equation}
If we now act with $\rot$ on the relations \eref{clc-points}, and use that
for all points $P$ with indices $1$ and $2$ we have $\rot(P_1)=P_2$ and
$\rot(P_2)=P_1$, then we find that
\begin{equation}
  \label{riso-points}
  A_1 = \riso_1(C_2), \quad
  B_1 = \riso_1(D_2), \qquad
  A_2 = \riso_2(C_1), \quad
  B_2 = \riso_2(D_1).
\end{equation}
It follows that the left moving light rays $A_1C_2$ and $B_1D_2$,
extended beyond these points, are fixed lines of $\riso_1$. Similarly,
the extended light rays $A_2C_1$ and $B_2D_1$ are fixed lines of
$\riso_2$. In other words, the light rays which, after the cutting and
gluing procedure, form the closed light rays on $\scri$ are those left
moving light rays on $\scri_0$, which are fixed lines of the combined
isometries $\riso_1$ and $\riso_2$.

The conformal boundary $\scri$ of $\ads$ splits into three open subsets
$\bnd_-$, $\sng$, and $\bnd_+$, separated by the two closed light rays,
as indicated in \fref{ctc}. In the shaded region
$\sng=\sng_1\cup\sng_2$, the time gained when passing over the cuts is
even bigger than $\pi/2$. This has two consequences.  First of all,
there are also closed timelike curves in that region. One such curve is
shown in the figure.  Moreover, consider a left moving light ray in
$\sng$.  After each winding, it appears a little bit earlier, so that
effectively it travels backwards in time.  Asymptotically, it approaches
the lower closed light ray. A timelike curve can have the same
behaviour. It follows that the causal structure of $\sng$ is completely
degenerate.  Every two points in $\sng$ can be connected by a future
pointing timelike curve.

\subsubsection*{Closed timelike curves}
The fact that there are closed timelike curves on $\scri$ implies that
such curves also exist in $\ads$. Consider, for example, the closed
timelike curve shown in \fref{ctc}, and shift it, by a continuous
deformation, away from the boundary into the interior of the cylinder.
For a sufficiently small deformation parameter it will still be
timelike, and thus it becomes a closed timelike curve in $\ads$. The
natural question that arises is which part of $\ads$ is filled with
closed timelike curves. The boundary of that region is called the
\emph{chronology horizon}. As we shall see, and also according to some
general theorems \cite{Cutler,Thorne}, the chronology horizon is a
lightlike surface that consists of pieces of null planes.

To see where it is located, let us first summarize some properties of
closed timelike curves. A quite general feature is that, within a
\emph{connected} region filled with closed timelike curves, the causal
structure is always completely degenerate. We saw this already in the
case of the region $\sng$ on $\scri$. In general, let $\ctc$ be a
connected subset of some spacetime, such that through each point
$P\in\ctc$ there passes a closed timelike curve. For each point
$P\in\ctc$, we define the subset $\ctc_P\subset\ctc$ containing all
points $Q\in\ctc$ such that there is a closed timelike curve passing
through $P$ and $Q$. It follows immediately from the definition that
$P\in\ctc_P$, and given two points $P$ and $Q$, then $\ctc_P$ and
$\ctc_Q$ are either equal or disjoint.

Moreover, $\ctc_P$ is always an open subset of $\ctc$. If a point $Q$ is
connected to $P$ along a closed timelike curve, then this is also true
for some neighbourhood of $Q$. For example, take the intersection of the
future light cone of some point on the curve shortly before $Q$ with the
past light cone of some other point shortly after $Q$. This is a
non-empty open set containing $Q$ and contained in $\ctc_P$. From all
this it follows that $\ctc=\bigcup_{P\in\ctc}\ctc_P$ is a decomposition
of $\ctc$ into non-empty, disjoint, open subsets. As $\ctc$ is assumed
to be connected, this is only possible if $\ctc_P=\ctc$ for all
$P\in\ctc$. Hence, every two points in $\ctc$ are connected along a
closed timelike curve. Or, equivalently, for every two points
$P,Q\in\ctc$ there is a future pointing timelike curve from $P$ to $Q$.

What does this imply for the chronology horizon in our spacetime
manifold? In the appendix we show that the subset $\ctc\subset\ads$,
which is defined to be the union of all closed timelike curves, is
indeed connected. It then follows that $\ctc$ is exactly that region of
$\ads$ which is \emph{causally connected} to $\sng\subset\scri$. By
causally connected we mean the set of all points $P\in\ads$ such that a
signal can be sent from $\sng$ to $P$ and from $P$ to $\sng$. As $\sng$
is an open subset of $\scri$, it is thereby sufficient to consider
signals traveling on timelike curves. The proof splits into two parts.
First we have to show that, given a point $P\in\ads$ which is causally
connected to $\sng$, then there is a closed timelike curve passing
through $P$.  And secondly, we have to show that every point on a closed
timelike curve in $\ads$ is causally connected to $\sng$.

The first part is quite easy. By assumption, there is a future pointing
timelike curve from some point $O\in\sng$ to $P$ and from there to
another point $Q\in\sng$. On the other hand, we know that every two
points on $\sng$ are connected by a future pointing timelike curve in
$\sng$. Hence, there exists a closed timelike curve through the points
$OPQ$. Parts of this curve lie on $\scri$. Using the same argument as
above, we can deform it such that it becomes a closed timelike curve in
$\ads$, still passing through $P$. Hence, there is indeed a closed
timelike curve in $\ads$ passing through every point which is causally
connected to $\sng$. The reversed statement follows from the previously
discussed general feature of closed timelike curves, applied to the
subset $\ctc\cup\sng$ of the compactified cylinder $\ads\cup\scri$.
Since this is a connected region filled with closed timelike curves, it
follows that every point $P\in\ctc$ is causally connected to $\sng$.

Finally, a more special feature of closed timelike curves in $\ads$
gives us some qualitative information about the shape of $\ctc$. Every
closed timelike curve in $\ads$ passes alternatingly over the two cut
surfaces, and it never hits the world lines of the particles. The first
property follows from the fact that there is no timelike curve in
anti-de~Sitter space which intersects a null plane twice. Hence, there
cannot be any closed timelike curve in $\ads$ passing over the same cut
surface twice without passing over the other in between. Moreover, if a
closed timelike curve intersects with a world line, then, by the same
argument, it cannot pass over the cut surface attached to \emph{this}
world line before or after the intersection. But then it has to pass
twice over the \emph{other} cut surface without a time jump in between,
which is also impossible.  All together, it is therefore reasonable to
expect that $\ctc$ is some torus like region surrounding the particles.
This is also what it looks like in the flat Gott universe.

\subsubsection*{The covering of $\sng$}
Let us now explicitly find the future and the past of $\sng$. Consider a
point $P\in\ads$, for example in the neighbourhood of the rectangle
$\sng_1$ in \fref{ctc}. That is, $P$ lies in the \emph{interior} of the
cylinder, but close enough to the boundary so that it makes sense to say
that $P$ lies \emph{above} the surface $\srf_1$ and \emph{below}
$\srf_2$. We can then ask whether a signal can be sent from $\sng$ to
$P$. Or, in other words, whether $\sng$ can be seen from $P$. If $P$
lies in the future of the point $A_1$ on the boundary, then this is
certainly possible. Now assume that it is not possible to reach $P$ from
$A_1$.  Then it might still be possible to reach it from some point in
the rectangle $\sng_2$, by passing through the cut surface $\srf_1$. As
seen from $P$, the rectangle $\sng_2$ behind the cut surface $\srf_1$
appears to be at the location
$\iso_1(\sng_2)=\iso_1(\rot(\sng_1))=\riso_1(\sng_1)$.

This is the first dotted rectangle beyond the points $A_1$ and $B_1$ in
\fref{ctc}.  For a signal to be sent from $\sng_2$ through $\srf_1$ to
$P$, it is sufficient for $P$ to lie in the future of the lower corner
of \emph{this} rectangle. We can obviously proceed this way. A signal
from $\sng$ to $P$ can also be sent from $\sng_1$, by passing through
the cut surface $\srf_2$ and then through $\srf_1$. This is possible if
$P$ lies in the future of the lower corner of the rectangle
$\iso_1(\iso_2(\sng_1))=\riso_1^2(\sng_1)$. This is the apparent
position of $\sng_1$ as seen from $P$, by looking through $\srf_1$ and
then through $\srf_2$. Allowing the signal to wind around arbitrarily
many times, we find that it is sufficient for the point $P$ in the
interior of the cylinder to lie in the future of the lower corner of at
least one of the rectangles $\riso_1^n(\sng_1)$ on the boundary, where
$n$ is a non-negative integer. It then also lies in the future of all
such rectangles with larger $n$.

Vice versa, we can ask the question as to whether a signal can be sent
from $P$ to $\sng$. If the point $P$ inside the cylinder lies in the
past of $D_2$ on the boundary, then the signal can be sent directly to
$\sng_1$. If we allow the signal to pass once over $\srf_2$, it is
sufficient for $P$ to lie in the past of the upper corner of the
rectangle $\iso_2^{-1}(\sng_2)=\riso_1^{-1}(\sng_1)$. This is the
apparent position of $\sng_2$ when looking from $P$ through the cut
surface $\srf_2$. To send a signal from $P$ through $\srf_2$ and
$\srf_1$ to $\sng_1$, $P$ has to lie in the past of the upper corner of
$\iso_2^{-1}(\iso_1^{-1}(\sng_2))=\riso_1^{-2}(\sng_1)$, and so on. For
a signal to be sent from $P$ to $\sng$ along any possible path, the
condition is that $P$ has to lie in the past of the upper corner of at
least one of the rectangles $\riso_1^{-n}(\sng_1)$, again for some
non-negative integer $n$.

All together, we can say that the \emph{apparent shape} of the strip
$\sng$ on the boundary, as seen by an observer located above $\srf_1$
and below $\srf_2$ in the interior of the cylinder, is represented by
the union of all rectangles $\riso_1^n(\sng_1)$, $n\in\ZZ$. Similarly,
we can make the same construction for an observer in the neighbourhood
of $\sng_2$. For her, the strip appears to look like the union of the
rectangles $\riso_2^n(\sng_2)$. Let us denote this by
\begin{equation}
  \label{cov-sng}
  \cov\sng_1 = \bigcup_{n\in\ZZ} \riso_1^n(\sng_1), \qquad
  \cov\sng_2 = \bigcup_{n\in\ZZ} \riso_2^n(\sng_2).
\end{equation}
Both $\cov\sng_1$ and $\cov\sng_2$, considered as subsets of $\scri_0$,
represent the \emph{covering space} of the strip $\sng$ on $\scri$. In
fact, the quotient spaces $\cov\sng_1/\riso_1^2$ and
$\cov\sng_2/\riso_2^2$ are, by construction, isometric to $\sng$. What
does this covering space look like? It is a strip bounded by the
extended left moving light rays $A_1C_2$ and $B_1D_2$, respectively
$A_2C_1$ and $B_2D_1$. As these are fixed lines of $\riso_1$,
respectively $\riso_2$, all the rectangles share them as their upper
right and lower left edges.  Furthermore, the rectangles fit together
and form a continuous strip, because the lower right edge of each
rectangle coincides with the upper left edge of the next one.

The crucial question is whether the strips are infinitely long or not.
This depends on whether the isometries $\riso_1$ and $\riso_2$ have
fixed points on $\scri_0$ or not. If they do, then the corners of the
rectangles $\riso_1^n(\sng_1)$ and $\riso_2^n(\sng_2)$ converge to the
first fixed points of $\riso_1$ and $\riso_2$ beyond the cuts. The
strips are then of finite size.  The most convenient way to find out
whether there are any fixed points is to look for fixed right moving
light rays of $\riso_1$ and $\riso_2$.  The fixed points are then at the
intersections with the left moving ones, which we already know.
Moreover, the first fixed right moving light rays beyond the cuts are
then the upper left, respectively the lower right edges of the
rectangles $\cov\sng_1$ and $\cov\sng_2$.
\begin{figure}[t]
  \begin{center}
    \epsfbox{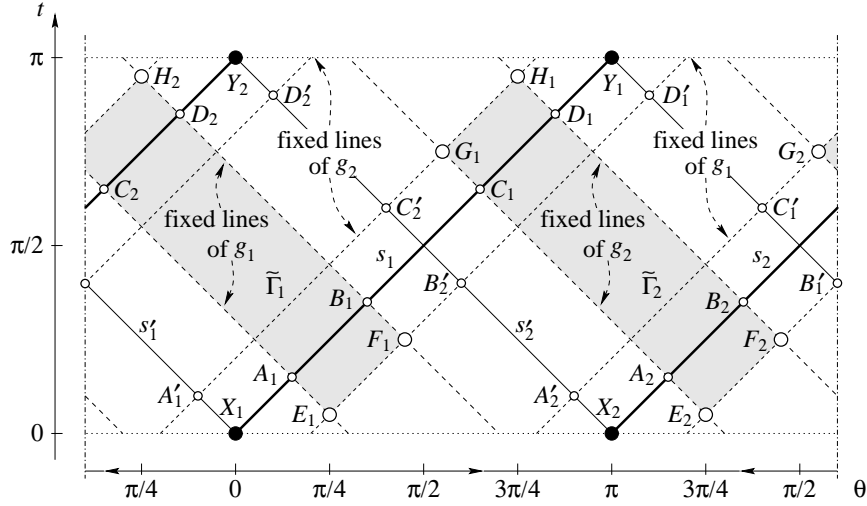}
    \caption{The construction of the fixed points $E_1$, $F_1$, $G_2$,
      $H_2$ of $\riso_1$, and $E_2$, $F_2$, $G_1$, $H_1$ of $\riso_2$.
      The solid lines are the fixed light rays of $\iso_1$ and $\iso_2$,
      and the dashed lines are those of $\riso_1$ and $\riso_2$.  The
      shaded rectangles $\cov\sng_1$ and $\cov\sng_2$ represent the
      apparent shape of the region of closed timelike curves on $\scri$,
      as seen by an observer in the interior of the cylinder, sitting in
      the neighbourhood of the rectangles $\sng_1$ and $\sng_2$ in
      \fref{ctc}.}
    \label{fix}
  \end{center}
  \hrule
\end{figure}

To find the fixed right moving light rays, we repeat the construction of
the left moving ones, with right and left interchanged. How did we find
the left moving ones? We started from the cut lines $\srf_1$ and
$\srf_2$ on $\scri_0$. These are the \emph{right} moving light rays
connecting the points $X$ and $Y$. They are once again shown as bold
lines in \fref{fix}. On these lines, we found the points $A$, $B$, $C$,
$D$, determined by \eref{clc-times}, such that $\iso(C)=A$, $\iso(D)=B$,
$t_C=t_A+\pi/2$ and $t_D=t_B+\pi/2$, with all indices equal to either
$1$ or $2$. From this we concluded that the \emph{left} moving light
rays passing through $A_1C_2$ and $B_1D_2$ are fixed lines of $\riso_1$,
and those passing through $A_2C_1$ and $B_2D_1$ are fixed lines of
$\riso_2$.

Now, we repeat this construction starting from the \emph{left} moving
light rays $\srf'_1$ and $\srf'_2$ connecting the points $X$ and $Y$.
These are the thin solid lines in \fref{fix}. We know already how the
isometries $\iso_1$ and $\iso_2$ act on them. They are the opposite
intersections of the null planes with the boundary of the cylinder. The
action of $\iso_1$ on $\srf'_1$ and that of $\iso_2$ on $\srf'_2$ is
given by \eref{scri-iso-2}. A point with time coordinate $t'_-$ is
mapped onto the one with time coordinate $t'_+$, where
\begin{equation}
  \label{fix-jump}
  \cot t'_+ - \cot t'_- =  - 2 \, \expo{\dis} \, \tan\erg . 
\end{equation}
In contrast to \eref{clc-jump}, this is a shift forward in time, and the
amount of shift is larger. We can always find four points $A'$, $B'$,
$C'$, and $D'$ on each cut, such that
\begin{equation}
  \label{fix-points}
  C'_1=\iso_1(A'_1), \quad
  D'_1=\iso_1(B'_1), \qquad
  C'_2=\iso_2(A'_2), \quad
  D'_2=\iso_2(B'_2),
\end{equation}
and with the time coordinates satisfying
\begin{equation}
  \label{fix-dif}
  t_{C'}=t_{A'}+\pi/2, \qquad t_{D'}=t_{B'}+\pi/2.
\end{equation}
The solution is similar to \eref{clc-times}, we only have to replace
$\mu$ by the second parameter $\nu$, as defined in \eref{mu-nu},
\begin{equation}
  \label{fix-times}
  \cot t_{A'} = \expo{\nu/2}, \qquad
  \cot t_{B'} = \expo{-\nu/2}, \qquad
  \cot t_{C'} = -\expo{-\nu/2}, \qquad
  \cot t_{D'} = -\expo{\nu/2}.
\end{equation}
Note that $\nu$ is related to $\mu$ by interchanging $\dis$ with
$-\dis$. Up to an overall sign, this is also the relation between
\eref{fix-jump} and \eref{clc-jump}. The sign is taken care of by
interchanging the points with their images. We have $A=\iso(C)$ and
$B=\iso(D)$, but $C'=\iso(A')$ and $D'=\iso(B')$. As a consequence,
the \emph{right} moving light rays passing though $A'_2C'_1$ and
$B'_2D'_1$ are now the fixed lines of $\riso_1$, and those passing
through $A'_1C'_2$ and $B'_1D'_2$ are the fixed lines of $\riso_2$.

Finally, we find the relevant fixed points of $\riso_1$ to be those
denoted by $E_1$, $F_1$, $G_2$, $H_2$ in \fref{fix}. These are the
limits to which the corners of the rectangles $\riso_1^n(\sng_1)$
converge as $n\to\pm\infty$. The union $\cov\sng_1$ of all these
rectangles is again a rectangle. Similarly, $\cov\sng_2$ is the
rectangle with corners at $E_2$, $F_2$, $G_1$, $H_1$. These are the
first fixed points of $\riso_2$ beyond the cuts. For the time
coordinates of the fixed points we have the relations
\begin{eqnarray}
  \label{riso-times}
  t_E &=& t_A - t_{A'} = t_C - t_{C'}, \qquad
  t_F = t_B - t_{A'} = t_D - t_{C'}, \nwl
  t_G &=& t_C + t_{A'} = t_A + t_{C'}, \qquad
  t_H = t_D + t_{A'} = t_B + t_{C'}.
\end{eqnarray}
Note that all these points lie in the time interval $0<t<\pi$, which
follows from \eref{clc-times} and \eref{fix-times}, under the condition
that $\mu<\nu$, which holds because of $\dis>0$.

\subsubsection*{The chronology horizon}
It is now straightforward to find the region of $\ads$ that is causally
connected to $\sng$, and thus the location of the chronology horizon.
We have to evolve the past light cone $\chr_+$ from the apparent
position of the \emph{last point} on $\sng$, respectively the future
light cone $\chr_-$ from the apparent position of the \emph{first point}
on $\sng$. Depending on which sides of the cut surfaces we are on, these
are either the null planes emerging from $E_1$ and arriving at $H_2$, or
those emerging from $E_2$ and and arriving $H_1$. The region that is
enclosed by these null planes and the cut surfaces is the subset
$\ctc\subset\ads$ that is filled by the closed timelike curves.
\begin{figure}[t]
  \begin{center}
    \epsfbox{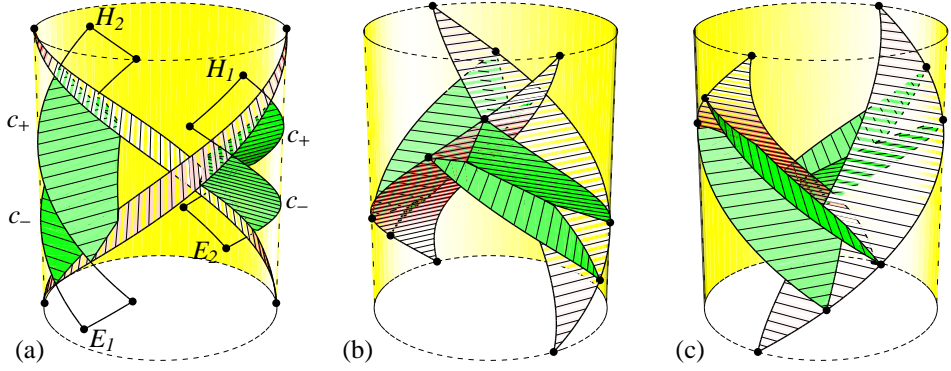}
    \caption{The chronology horizon consists of two null planes in
      $\ads$, denoted by $\chr_-$ and $\chr_+$. Each of them is defined
      piecewise, by gluing together appropriate parts of null planes in
      $\ads_0$ (a). For $\chr_-$, they originate from the points
      $E_{1,2}$, and for $\chr_+$ their destinations are the points
      $H_{1,2}$. The horizon encloses the region $\ctc$ of closed
      timelike curves, which has the shape of a triangular torus (b),
      that winds around the particles without touching them (c).}
    \label{chr}
  \end{center}
  \hrule
\end{figure}

In \fref{chr}(a), the shaded rectangles from \fref{fix}, with the fixed
points of the isometries $\riso$ in the corners, are shown on the
boundary of the cylinder. Evolving the null planes backwards from $H_1$
and $H_2$, and taking only the pieces that lie on the correct sides of
the cut surfaces, we get the upper null plane $\chr_+$ of the chronology
horizon. What is important to note is that the two pieces fit together
along the cut surfaces, forming a single null plane in $\ads$.
Consider, for example, the neighbourhood of the cut surface $\srf_1$.
This is the one in the front of \fref{chr}(a). Below $\srf_1$, the
horizon is represented by a piece of the null plane with destination
$H_1$. A piece of this plane immediately below $\srf_1$ can be seen on
the right side in \fref{chr}(a).

It continues above $\srf_1$, as a piece of the null plane with
destination $H_2$. This piece of the horizon $\chr_+$ can be seen almost
completely on the left side of \fref{chr}(a). For these two pieces to
form a single null plane in $\ads$, without a kink at the intersection
with the cut surface $\srf_1$, the lower part must be related to the
upper part by the transition function $\iso_1$.  This is in fact the
case, because for the points $H_1$ and $H_2$ defining the null planes we
have $\iso_1(H_1)=\riso_1(\rot(H_1))=\riso_1(H_2)=H_2$.  The last
equation holds because, by definition, $H_2$ is a fixed point of
$\riso_1$.  Similarly, one shows that $\chr_+$ is smooth at the
intersection with $\srf_2$, and thus a null plane in $\ads$.  The lower
part $\chr_-$ of the chronology horizon is defined in the same way.

Instead of the null planes with destinations $H$, we take appropriate
pieces of those with origins $E$. They also form a single null plane in
$\ads$. It intersects with $\chr_+$ along a special curve in $\ads$.
Being the intersection of two null planes, it is a spacelike geodesic.
On the other hand, it is a closed loop. It is thus a spacelike closed
geodesic, and in fact the only closed geodesic in $\ads$. This is also
where the horizon ends, because the points beyond that intersection are
no longer causally connected to $\sng$. So, we can say that the
chronology horizon extends from spatial infinity to a closed spacelike
geodesic that winds around the particles. Again, the chronology horizon
of the flat Gott universe has exactly the same behaviour \cite{Cutler}.

\section{The BTZ Wormhole}
\label{wormhole}
Now we are going to change our point of view. Instead of considering the
spacetime $\ads$ as a generalized Gott universe, we would like to think
of it as a three dimensional black hole of the BTZ type. To do this, we
have to interpret the region containing the closed timelike curves as a
\emph{singularity}. This is the typical procedure to construct three
dimensional black holes. For example, to get a matter free black hole,
one starts from empty anti-de~Sitter space and divides it by the action
of some discrete isometry group \cite{BTZ,BHTZ,ViBrill}.  The quotient
space typically contains closed timelike curves. To get a well defined
causal structure, one takes away a subset through which all closed
timelike curves have to pass, and interprets the boundary of that subset
as a singularity.

We can do the same here. The question is then, which part of the region
$\ctc$ behind the chronology horizon do we have to take away in order to
remove all closed timelike curves? In the case of the quotient space
construction, with the isometries defined by one or several Killing
vectors, one usually takes away the region where the Killing vectors are
timelike. This region is obviously filled with closed timelike curves,
namely the flow lines of the Killing vectors. On the other hand, one
can show that every closed timelike curve must pass through this region.
However, even in the simplest case of only one Killing vector, the
singular subset to be removed is not \emph{minimal}, in the sense that
there is no smaller set through which all closed timelike curves pass.

In fact, such a minimal subset does not exist, and therefore the only
motivation to choose the one defined by the timelike Killing vectors is
to preserve the symmetries of the spacetime. If we want make a similar
construction here, we first have to deal with the problem that our
spacetime is not the quotient space of anti-de~Sitter space with respect
to some discrete isometry group. However, a part of it turns out to have
a rotational symmetry. There exists a unique rotational Killing field
$\kllr$, with the property that $\expo{2\pi\kllr}=\id$. In other words,
at least a part of our spacetime looks exactly like a part of a rotating
vacuum BTZ wormhole, which is obtained as a quotient space of
anti-de~Sitter space with respect to the isometry generated by $\kllr$.

Of course, this Killing field cannot be defined globally on $\ads$,
because obviously our spacetime does not possess a continuous rotational
symmetry. But it can be defined within a sufficiently large subset,
which includes the region of closed timelike curves. An explicit
construction of $\kllr$ and its maximal support is given in the
appendix. All we need to know here is that, using the definition
\eref{kll-def}, it can be written as 
\begin{equation}
  \label{kllr}
  \kllr(\xx) = \xx \, \nn - \mm \, \xx, \txt{where}
       \mm^2 = \mu^2 \, \one , \quad \nn^2 = \nu^2 \, \one .
\end{equation}
Hence, $\mm,\nn\in\alg$ are two spacelike Minkowski vectors with lengths
$\mu$ and $\nu$. Depending on which sides of the cut surfaces we are on,
they are explicitly given by one of the expressions in \eref{riso-log}.
The singular region can then be defined to be that subset of $\ads$
where $\kllr$ is timelike. According to what has been said above, if
follows that this region lies behind the chronology horizon, and that
every closed timelike curve has to pass through it.

If we take it away, the resulting spacetime has a well defined causal
structure. Its boundary is a naked singularity. It is the surface where
$\kllr$ is lightlike. Since $\kllr$ is also tangent to it, it must be a
timelike surface. Hence, timelike curves can end on or start from the
singularity. We are not going to compute the precise form of the
singularity. From the properties listed so far, we can derive all
relevant features of the truncated spacetime. For example, we can
immediately infer that, although it is of course no longer a chronology
horizon, the null planes $\chr_\pm$ still form a horizon. It is now the
boundary of the region that is causally connected to the singularity.
\begin{figure}[t]
  \begin{center}
    \epsfbox{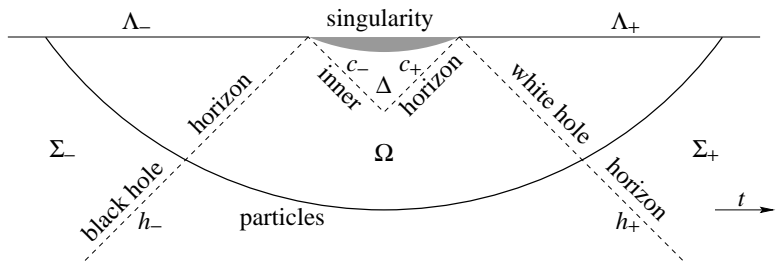}
    \caption{A schematic diagram of the wormhole spacetime. The dotted
      line below represents the central axis of the cylinder, the upper
      boundary is $\scri$, which splits into two disconnected components
      $\bnd_\pm$, representing the infinities of the two exterior
      regions $\ext_\pm$. The black hole horizon is the backward light
      cone of the last point on $\bnd_-$, and the white hole horizon is
      the forward light cone of the first point on $\bnd_+$. The inner
      horizon separates the interior $\wrm$ of the wormhole from the
      region $\ctc$ which is causally connected to the singularity.}
    \label{csl}
  \end{center}
  \hrule
\end{figure}

To see what happens to the causal structure of $\ads$ when the singular
region is taken away, it is again most convenient to look at $\scri$
first. On $\scri$, the region of closed timelike curves coincides with
the region where $\kllr$ is timelike, and thus with the singularity.
This is simply because it is two-dimensional, as one can easily convince
oneself. The two closed light rays on $\scri$ are in fact flow lines of
$\kllr$, and so is the closed timelike curve shown in \fref{ctc}. If we
take away the singular region $\sng$, then $\scri$ obviously falls apart
into two disconnected components $\bnd_-$ and $\bnd_+$. Hence, our new
universe has two distinct spacelike and lightlike infinities, and it is
reasonable to expect it to be a timelike wormhole that connects the two
infinities.

The resulting causal structure of the interior of $\ads$ is sketched in
\fref{csl}. There are two \emph{external} regions $\ext_-$ and $\ext_+$.
They are defined to be those parts of $\ads$ which are causally
connected to $\bnd_-$ and $\bnd_+$. In between, we have the interior of
the wormhole $\wrm$, with is causally connected to neither of them. It
is separated from $\ext_-$ by a black hole event horizon $\hor_-$, which
is the past light cone attached to the \emph{last point} on $\bnd_-$.
And vice versa, there is a white hole event horizon $\hor_+$ attached to
the \emph{first point} on $\bnd_+$, which separates the wormhole from
the second exterior region. Inside the wormhole, we have an additional
inner horizon $\chr_\pm$, which encloses the region that is causally
connected to the singularity.

The particles themselves pass through the wormhole, from one exterior
region to the other. An observer in $\ext_-$ sees the particles coming
from infinity, approaching each other, and falling into a black hole.
Another observer in $\ext_+$ sees them falling out of a white hole,
separating, and disappearing to spatial infinity. But not only the
lightlike world lines of the particles pass the wormhole. There are also
timelike geodesics passing through, for example the central axis of the
cylinder. The wormhole obviously allows an observer to pass through
without hitting the singularity, and even without crossing the inner
horizon. Somewhat sloppy speaking, the passage through the wormhole is
quite safe, if one sticks to the region in between or close to the
particles.

This is not so in the case of the standard version of the rotating BTZ
wormhole, with no matter inside. Apart from the fact that this consists
of a whole series of exterior regions connected by timelike wormholes,
the crucial difference is that there it is not possible to pass from one
exterior region to the next without crossing, or at least touching the
inner horizon. The reason is that before the singular region is removed,
the closed timelike curves in the quotient space fill a subset that
separates the two exterior regions from each other. Hence, every curve
connecting two exterior regions necessarily has to cross the chronology
horizon. As this becomes the inner horizon when the singular region is
taken away, it follows that every observer that passes the vacuum
wormhole has to pass the inner horizon.

\subsubsection*{The event horizons}
To locate the two event horizons, we can apply the same method that we
already used to find the chronology horizon. First we look for the
apparent positions of the last point on $\bnd_-$ and the first point on
$\bnd_+$. Then we evolve the past, respectively the future light cones
from there. From \fref{fix}, it is quite obvious that the relevant
points are $G_2$ and $F_1$, for an observer located in the interior of
the cylinder above $\srf_1$ and below $\srf_2$, and $G_1$ and $F_2$ for
an observer above $\srf_2$ and below $\srf_1$. Using the same arguments
as before, one shows, for example, that a signal can be sent from a
point above $\srf_1$ and below $\srf_2$ to $\bnd_-$ if and only if it
lies in the past of $G_2$.
\begin{figure}[t]
  \begin{center}
    \epsfbox{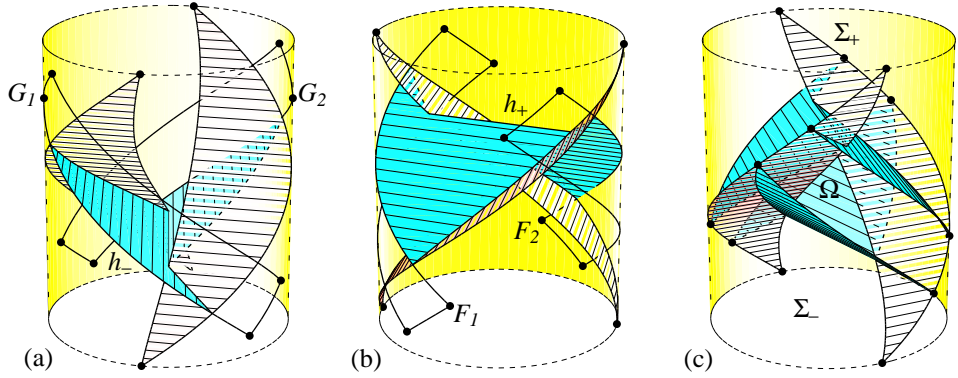}
    \caption{The black hole event horizon (a) is made out of two pieces
      of null planes, with destinations $G_1$ and $G_2$, forming a
      single null plane $\hor_-$ in $\ads$. It emerges from a cusp,
      located on the spacelike geodesic connecting the particles in the
      moment when they fall behind it. The white hole horizon (b)
      consists of pieces of the null planes with origin $F_1$ and $F_2$,
      forming a null plane $\hor_+$ in $\ads$. Together (c) they enclose
      the interior of the wormhole.}
    \label{hor}
  \end{center}
  \hrule
\end{figure}

The relevant pieces of the null planes with destination $G_1$ and $G_2$
are shown in \fref{hor}(a). They form the black hole event horizon
$\hor_-$. The subset below is the exterior region $\ext_-$. Again, the
horizon is a single null plane in $\ads$. The two pieces fit together
correctly along the cut surfaces, because the points $G_1$ and $G_2$ are
fixed points of the isometries $\riso_2$ and $\riso_1$. The same holds
for the white hole event horizon $\hor_+$, which consists of two pieces
of the null planes which originate from the points $F_1$ and $F_2$. This
is shown in \fref{hor}(b). Together they enclose the wormhole region
$\wrm$, as can be seen in \fref{hor}(c).

What is different to the inner horizon is that the event horizons reach
the particles. If we follow, for example, the null planes attached to
$G_1$ and $G_2$ backwards in time, then at some point they hit the
particles. Both particles are hit at the same time, and this is also the
moment when the two null planes intersect. This can be inferred from the
symmetry of our spacetime and the fact that the world lines are fixed
lines of $\iso_1$ and $\iso_2$. Hence, if one of the null planes hits a
particle, then the other does as well, and due to the symmetry under
rotations by $180$ degrees the same happens to the other particle at the
same time. It also follows that the two pieces of the horizon intersect
along the spacelike geodesic that connects the particles at this moment.

Note that this is a \emph{self} intersection of a single null plane in
$\ads$, which is due to the conical singularities on the world lines.  A
plane that intersects a conical singularity, which is not orthogonal to
the plane, necessarily has a kink on a line that points away from the
point of intersection. The horizons are exactly those null planes in
$\ads$, for which this kink is located on a geodesic connecting the
particles. We can think of this self intersection as a cusp from where
the horizon $\hor_-$ emerges. If we go further back in time, then we can
reach spatial infinity $\bnd_-$ by either sticking to one or the other
side of the null planes. We are then always in the region that is
causally connected to $\bnd_-$. We can say that the horizon $\hor_-$ is
created when the particles reach a critical distance, and at the same
time they fall behind it.

The same, but with the time direction reversed, happens in the second
exterior region $\ext_+$. An observer sees the particles emerging from a
white hole. At the same moment, the event horizon $\hor_+$ degenerates
to a spacelike geodesic connecting the particles, and then it
disappears. The time when the black hole horizon is formed can be easily
calculated. The two null planes of which it is made of are those that
arrive at the points $G_1$ and $G_2$ at $t=t_G$. As these are two
antipodal points on the boundary of the cylinder, the null planes
intersect in the middle of the cylinder at $t=t_G-\pi/2=t_A+t_{A'}$.
Similarly, one finds that the time when the white hole horizon
disappears is $t=t_F+\pi/2=t_D-t_{A'}$.

This can be used to compute the length $\len$ of the horizon. From the
moment it is created the length is constant. This is because, as we just
saw, the horizon is a null plane with no kinks other than the cusp from
where it emerges. Immediately after its creation, the length is twice
the distance between the particles. So, what we need to know is the
distance between the particles. The most convenient way to compute this
is to get back to the hyperbolic radial coordinate $\ch$ on
anti-de~Sitter space. According to \eref{ads-metric-1}, it measures the
physical distance of a point in $\ads_0$ from the central axis of the
cylinder at $\ch=0$. Due to the symmetry, the radial coordinate $\ch$ of
the points where the particles hit the horizon is the same for both
particles, and the horizon length is then $\ell=4\ch$.

We know how to transform the spherical coordinates back to the
hyperbolic ones. Equation \eref{ch-qh-ph} tells us that
\begin{equation}
  \cosh\Big(\frac{\ell}{4}\Big) = \frac{1}{\sin \qh \, \sin \ph},
\end{equation}
where $\qh$ and $\ph$ are the spherical coordinates of either particle
one or two at the moment when the horizon is created. Let us choose
particle one, whose world line $\prt_1$ is defined by \eref{p-lines}.
The coordinate $\ph$ is then given by $\cos\ph=\tanh\dis$, which implies
that $\sin\ph=1/\cosh\dis$, and $\qh$ is equal to $t=t_A+t_{A'}$.
Inserting the values for $t_A$ and $t_{A'}$ from \eref{clc-times} and
\eref{fix-times}, and expressing the parameter $\dis$ in terms of $\mu$
and $\nu$, as defined in \eref{mu-nu}, one arrives at the simple
relation
\begin{equation}
  \label{hor-len}
  \len = \mu + \nu .
\end{equation}
This agrees with the result that for a non-rotating black hole formed by
colliding particles, where $\dis=0$ and therefore $\mu=\nu$, the horizon
length is $\len=2\mu$ \cite{Matschull}. 

\subsubsection*{The angular velocity}
Not surprisingly, our wormhole also has a non-vanishing angular velocity.
Let us briefly remind ourselves how this is defined \cite{Wald}. As
already discussed in the beginning, for an axially symmetric black hole
one has a rotational Killing vector $\kllr$, which is uniquely defined
by the condition that $\expo{2\pi\kllr}=\id$. If the black hole is also
static, then there is a second Killing vector $\kllt$, which is
hypersurface orthogonal and generates time translations. It is also
orthogonal to $\kllr$, and equal to the horizon generating Killing
vector $\kllh$. What is meant by this is that $\kllh$ is the Killing
vector which is spacelike outside, lightlike on, and timelike inside the
black hole, and whose geodesic flow lines are the light rays that travel
along the horizon.

If the black hole is not static but still stationary, then the Killing
vector $\kllt$ is no longer hypersurface orthogonal, and it can no
longer be chosen to be orthogonal to $\kllr$. However, it is still
uniquely, up to rescaling, defined by the condition that $\kllr$ and
$\kllt$ should be orthogonal on $\scri$, or \emph{asymptotically
  orthogonal} in the sense of \eref{kll-ortho}. The horizon is still a
fixed surface of both rotations and time translations, and thus $\kllr$
and $\kllt$ are both tangent to it.  However, there is only one linear
combination,
\begin{equation}
  \label{ang-def}
  \kllh = \kllt + \ang \, \kllr,
\end{equation}
which is also tangent to the individual light rays spanning the horizon,
and this defines the angular velocity $\ang$ of the black hole. In
asymptotically flat spacetimes, $\kllt$ can be expressed in physical
units, so that $\ang$ gets the correct dimension $1/$time. In
anti-de~Sitter space, we can only refer to the coordinate time $t$. A
possible way to normalize $\kllt$ is to require that on $\scri$ its norm
is the same as that of the rotational Killing vector,
$\kllt^2=-\kllr^2$, with respect to the conformally transformed metric
\eref{conf-metric}. This is independent of the conformal factor, and we
get the same normalization as in empty anti-de~Sitter space, where
$\kllt=\del_t$ and $\kllr=\pm\del_\qh$.

Now, we have the expression \eref{kllr} for the rotational Killing
vector $\kllr$, and from \eref{kll-ortho} we know how to construct an
asymptotically orthogonal Killing vector $\kllt$, namely
\begin{equation}
  \label{ang-kll-rot}
  2\pi \, \kllr(\xx) = \xx \, \nn  -  \mm \, \xx \follows
  2\pi \, \kllt(\xx) = {} - \xx \, \nn  -  \mm \, \xx.
\end{equation}
Up to a sign, $\kllt$ is determined by the condition that
$\kllt^2=-\kllr^2$. The sign is chosen such that $\kllt$ is pointing
towards the future, with $\mm$ and $\nn$ given by \eref{riso-log}.
Inserting this into \eref{ang-def}, we get
\begin{equation}
  \label{ang-kll-hor}
    2\pi \, \kllh(\xx) = 
      \xx \, \nn \, (\ang-1) - (\ang+1) \, \mm \, \xx.
\end{equation}
What we have to find is that value of $\ang$ for which $\kllh$ is
tangent to the light rays on the horizon. This can be done without
explicitly calculating $\kllh$ on the horizon. We know that the horizon
is made out of pieces of null planes. We also know that these null
planes are fixed surfaces of the Killing vectors $\kllr$ and $\kllt$,
and hence of $\kllh$ for any given value of $\ang$. Among all the
possible linear combination of $\kllr$ and $\kllt$, we have to find that
one which also has the individual light rays as fixed lines.

In \sref{math} we saw that the condition \eref{kll-tng-con} for the
Killing vector $\kllh$ to have a family of fixed light rays is that the
Minkowski vectors appearing to the left and to the right of $\xx$ in the
definition \eref{ang-kll-hor} are either lightlike or spacelike and of
the same length. This leads to the condition that
\begin{equation}
  \label{ang-con}
  \nn^2 \, (\ang-1)^2 = \mm^2 (\ang+1)^2 \follows
  \nu \, (\ang - 1) \pm \mu \, (\ang+1) = 0.
\end{equation}
There are two solutions, namely 
\begin{equation}
  \label{ang-vel}
  \ang = \frac{\nu - \mu}{\nu + \mu} , \qquad
  \tilde\ang = \frac{\nu + \mu}{\nu - \mu}.
\end{equation}
Now, why do we get two solutions, and what is the correct one? First of
all, the condition only tells us that there is \emph{some} family of
fixed light rays. This need not be the event horizon, it can also be the
inner horizon, which is a second fixed null plane of both $\kllt$ and
$\kllr$. It is however easy to see which is the correct one.

For $\kllh$ to be timelike outside the horizon, and thus in particular
on $\scri$, where $\kllt$ and $\kllr$ are unit timelike and spacelike
vectors, we must have $|\ang|<1$. This implies that the first solution
$\ang$ must the correct one. It also gives the correct limit $\ang=0$
for $\mu=\nu$, which is the case when the particle collide and a static
black hole is formed. We shall discuss this in more detail below. The
second solution $\tilde\ang=1/\ang$ is the angular velocity of the inner
horizon. It goes to infinity in the static case, which is also quite
reasonable as we shall see.

Like every rotating black hole, our wormhole also has an ergosphere.
That is, a region of spacetime surrounding the event horizon where it is
impossible to stand still, with respect to the generating Killing vector
$\kllt$ of time translations. To see this, we have to show that $\kllt$
is spacelike on the event horizon, and thus already in a finite
neighbourhood of the horizon. Let us compute the norm of both $\kllt$
and $\kllr$ on the horizon. Using the formula \eref{kll-prod}, we find
that
\begin{equation}
  \label{kll-r-t-norm}
  4 \pi^2 \, \kllt^2 =
   \ft12\Trr{\nn^2 + \mm^2 + 2 \, \xx^{-1} \mm \, \xx \, \nn }, \quad
  4 \pi^2 \, \kllr^2 =
   \ft12\Trr{\nn^2 + \mm^2 - 2 \, \xx^{-1} \mm \, \xx \, \nn }.
\end{equation}
On the other hand, we know that the vector
\begin{equation}
  \label{kll-hor-sol}
  2 \pi \, \kllh = -\frac{2 \, \mu \, \xx \, \nn +
                          2 \, \nu \, \mm \, \xx}{\mu+\nu},
\end{equation}
obtained by inserting the correct value for $\ang$ into \eref{ang-def},
is lightlike on the event horizon, and thus
\begin{equation}
  \label{kll-hor-norm}
  \pi^2 \, (\mu+\nu)^2 \,\kllh^2 =
   \ft12\Trr{\mu^2 \nn^2 + \nu^2 \mm^2 
             + 2 \mu \nu \, \xx^{-1} \mm \, \xx \, \nn } = 0.
\end{equation}
Using that $\mm^2=\mu^2\one$ and $\nn=\nu^2\one$, this tells us that on
the event horizon we have $\Tr(\xx^{-1}\mm\xx\nn)=-2\mu\nu$. Inserting
this into \eref{kll-r-t-norm}, we find that
\begin{equation}
  \label{ergo}
  4 \pi^2 \, \kllt^2 =  (\nu - \mu)^2  , \qquad
  4 \pi^2 \, \kllr^2 =  (\nu + \mu)^2 .
\end{equation}
Hence, both $\kllr$ and $\kllt$ are spacelike on the event horizon. The
only exception is the static case, where $\mu=\nu$ and therefore $\kllt$
is lightlike. In this case, there is of course no ergosphere.

We can also define a generating Killing vector of the inner horizon,
which is obtained by just replacing $\ang$ with $\tilde\ang=1/\ang$, or
$\mu$ with $-\mu$. Repeating the same calculation gives that on the
inner horizon we have
\begin{equation}
  \label{sing}
  4 \pi^2 \, \kllt^2 =  (\nu + \mu)^2  , \qquad
  4 \pi^2 \, \kllr^2 =  (\nu - \mu)^2 .
\end{equation}
So, for generic values of $\mu$ and $\nu$, both $\kllr$ and $\kllt$ are
also spacelike on the inner horizon. In particular, $\kllr$ is not yet
lightlike, which means that the singularity is further behind the inner
horizon. The only exception is again the static case, where $\kllr$ is
lightlike on the inner horizon. In this case, it consists of a family of
closed lightlike geodesics, and this is already the singularity. This is
also the reason why the angular velocity $\tilde\ang$ of the inner
horizon diverges in the static case, whereas the angular velocity of the
event horizon goes to zero. The generating Killing vector of the inner
horizon coincides with $\kllr$, whereas the generating Killing vector of
the event horizon is $\kllt$.
 
\subsubsection*{Extremal and static black holes}
To complete the discussing of the wormhole spacetime, let us consider
some limits. We have two parameters $0<\mu<\nu$, which we can think of
as specifying the properties of the wormhole. They determine the horizon
length \eref{hor-len} and the angular velocity \eref{ang-vel}. They are
related to the parameters $\dis$ and $\erg$ by \eref{mu-nu}, describing
the relative motion of the particles. There are two limits we can take.
For $0<\mu=\nu$, the angular velocity of the black hole vanishes. In
this case we have $\dis=0$, which means that the particles collide. The
other special situation is $0=\mu<\nu$. This is the case when the
relevant function of the energy and the distance of the particles has
just reached the threshold, $\expo\dis=\tan\erg$. And finally, we can
also consider the very special situation $0=\mu=\nu$. 

Let us consider the second case first. For $0=\mu<\nu$, we have an
\emph{extremal} black hole. From \eref{clc-times} we infer that
$t_A=t_B$ and $t_C=t_D$, which means that the region $\sng$ of closed
timelike curves on $\scri$, as shown in \fref{ctc}, is degenerate to a
single closed light ray. There are no closed timelike curves at all on
$\scri$, and consequently also no closed timelike curves in $\ads$. This
also follows from the condition \eref{ctc-con-fac} for closed timelike
curves derived in the appendix. If we take the limit $\mu\to0$ with
fixed $\nu>0$, then the chronology horizon of the Gott universe shrinks
and finally disappears at $\scri$. But we can still consider the
remaining closed light rays on $\scri$ as a singularity, splitting
$\scri$ into two disconnected infinities $\bnd_-$ and $\bnd_+$.
\begin{figure}[t]
  \begin{center}
    \epsfbox{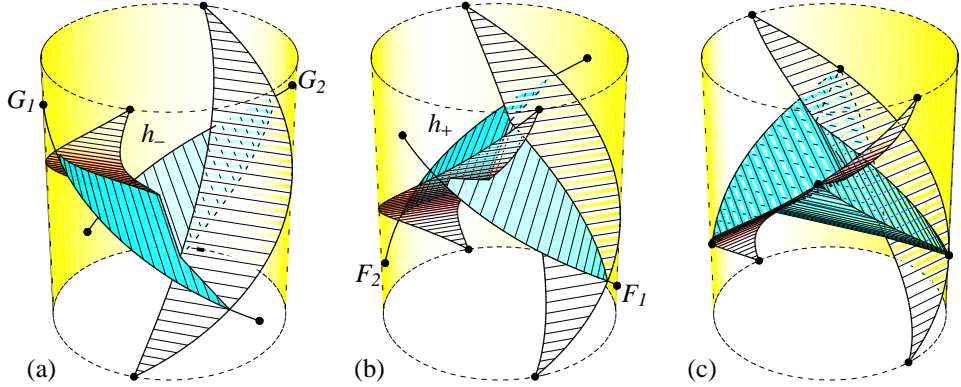}
    \caption{The extremal, or lightlike wormhole with $0=\mu<\nu$. The
      rectangles of \fref{fix} on the boundary are degenerate. There are
      no closed timelike curves, and the singularity consists of a
      single closed light ray on $\scri$. There is no inner horizon, but
      the event horizons $\hor_\pm$ are still there, consisting of
      pieces of null planes attached to the points $G=H$, and $E=F$.}
    \label{ext}
  \end{center}
  \hrule
\end{figure}

Evolving the past and future light cones from the apparent positions of
the last point on $\bnd_-$ and the first point on $\bnd_+$, we find the
black hole and the white hole horizon, enclosing the interior of the
wormhole. As indicated in \fref{ext}, the rectangles $\cov\sng$
constructed in \fref{fix} are also degenerate, and the origins and
destinations of the event horizons $F$ and $G$ now coincide
with the points $E$ and $H$. As a consequence, the event
horizons meet at the boundary of the cylinder, and there is no room any
more for an inner horizon and a singularity inside $\ads$. The wormhole
is infinitely large, because the singularity is infinitely far away. It
has also reached its maximal angular velocity $\ang=1$.

For $0<\mu=\nu$ we have the opposite situation. The angular velocity
vanishes, and we expect the wormhole to be static. In this case, we also
have $\dis=0$, which means that the particles collide at $t=\pi/2$ in
the centre of the cylinder. This process has already been studied in
\cite{Matschull}, with the result that two colliding particles form a
static black hole, but not a wormhole. Indeed, if we look at the
equations \eref{riso-times}, and take into account that for $\mu=\nu$ we
have $t_P=t_{P'}$ for all the points $P=A,B,C,D$ on the cuts, then we
find that $t_E=0$ and $t_H=\pi$. These are the origins and destinations
of the null planes that form the chronology horizon.
\begin{figure}[t]
  \begin{center}
    \epsfbox{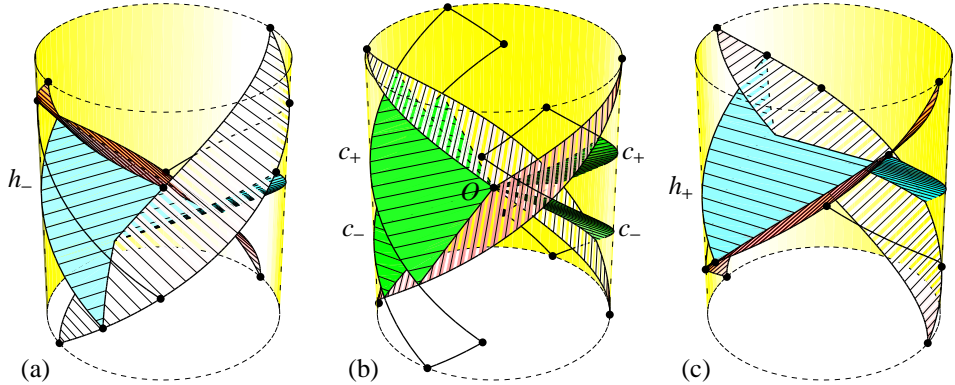}
    \caption{A static black hole formed by colliding particles with
      $0<\mu=\nu$. The chronology horizon reaches the world lines in the
      moment of collision $O$. When interpreted as a Gott universe, this
      spacetime is geodesically incomplete. When the singular region,
      which coincides with the interior of the chronology horizon, is
      removed, then the wormhole closes and the spacetime falls apart
      into a black hole and a white hole.}
    \label{stc}
  \end{center}
  \hrule
\end{figure}

If their distance in time is $\pi$, then the null planes emerging from
there meet at $t=\pi/2$ in the centre of the cylinder. But this is also
the point $O$ where the particles collide, as shown in \fref{stc}(b).
Hence, what happens in the special case of colliding particles is that
in the moment of collision they also hit the chronology, or inner
horizon. Even worse, this is already the singularity, which has been
defined to be the surface where $\kllr$ is lightlike. On the other hand,
we have seen that in the static case the angular velocity of the inner
horizon in infinite, which means that $\kllr$ is also the generating
Killing vector of the inner horizon. Hence, the inner horizon coincides
with the singularity and it touches the particles in the moment of
collision.

This agrees very nicely with the result of \cite{Matschull}, where it
has been shown that the only way to avoid closed timelike curves after
the collision of the particles is to assume that the particles stick
together and form a tachyonic object, which is interpreted as a future
singularity of a black hole. Here we have the same situation. Every
timelike curve that enters the event horizon necessarily ends on the
inner horizon, and thus on the singularity. Moreover, the singularity
also emerges, in a sense, from the point of collision of the two
particles. If we take away the singular region, then the spacetime falls
apart into two disconnected components, a black hole and its white hole
counterpart. So, we agree with the previous result, and we also find the
same horizon length of $\len=2\mu$.

We can think of the generalized Gott universe, with the closed timelike
curves not being removed, as a possible extension of the static black
hole \emph{beyond} the singularity. Such an extension necessarily
contains closed timelike curves. It was the ADM like formulation used in
the previous work to describe the colliding particles, which prevented
us from seeing how such an extension looks like. Now we have an explicit
representation in form of a generalized Gott universe. There remains,
however, one somewhat peculiar and rather unexpected feature of the
spacetime $\ads$ with $\mu=\nu$. Obviously, although the infalling
particles have no angular momentum, the spacetime shown in \fref{stc}
has an inherent orientation.

The closed timelike curves wind around the particles in one direction
but not in the other. This suggests that the extension of the static
black hole is not unique. There must be at least a second one, with
opposite orientation. We can also see this when we look more closely at
the scattering process of the two particles. When they hit each other,
it is not clear in which direction they should continue afterwards. Here
we assumed that the particles move on as if they had passed each other
on the right. We can also take the limit from the left. Then we obtain
an alternative extension of the static black hole, with the opposite
orientation. The third possibility is that the particles stick together.
This is the only way to avoid the closed timelike curves
\cite{Matschull}.

If there are several possible generalized Gott universes, which are all
extensions of the static black hole, then neither of them can be
geodesically complete. So far, we haven't considered this problem at
all. Actually, all our spacetimes might be geodesically incomplete. That
is, before the singular region is removed, since afterwards they are
incomplete anyway. If this is the case, then we probably miss some
interesting features. There might be a second singularity, for example.
Fortunately, it turns out that all universe constructed here are
geodesically complete, expect for those with $\mu=\nu$. The proof is
given in the appendix. In the static case, there are geodesics that wind
around the particles infinitely many times, such that with each winding
their affine length decreases. As a result, the total length of such a
geodesic converges.
\begin{figure}[t]
  \begin{center}
    \epsfbox{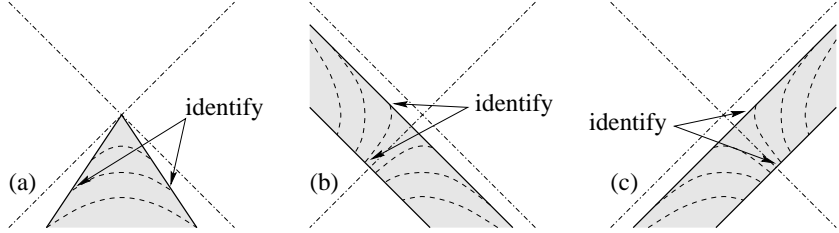}
    \caption{The Misner universe (a) is a wedge in the past of the origin
      of $1+1$ dimensional Minkowski space, with the faces identified by
      the action of a boost. It has a big crunch like singularity at the
      origin. It is isometric to the quotient space of the interior of
      the past light cone with respect to the given boost. There are two
      possible extensions, either including the left (b) or the right
      (c) spacelike region and then taking the quotient. Both contain
      closed timelike curves, and both are still incomplete. To get a
      geodesically complete spacetime, one has to include the whole
      Minkowski space, but then the quotient is no longer a Hausdorff
      space.}
    \label{msn}
  \end{center}
  \hrule
\end{figure}

The problem is of the same type as in the well known case of the
\emph{Misner universe} \cite{Misner}, which is sketched in \fref{msn}.
It is a simple two or higher dimensional locally flat and symmetric
spacetime with a singularity in the future, similar to that of a static
BTZ black hole. It can be extended beyond the singularity in two
different ways. Both contain closed timelike curves, and both have an
inherent orientation. The closed timelike curves wind either in one or
the other direction. Both extensions are still geodesically incomplete.
It is not possible to construct a geodesically complete spacetime from
the Misner universe which is at the same time a proper manifold. The
same is happening here as well. There are several possible extensions of
the static black hole, but none of them is geodesically complete.

Finally, let us also consider the extremal static case $0=\mu=\nu$. This
has also been discussed in \cite{Matschull}. At first sight, there seems
to be a contradiction. For the extremal wormhole we found that there is
no inner horizon and the singularity is infinitely far away. For the
static one it even reaches the particles. The solution is that in the
extremal static case there are indeed no closed timelike curves, but
there is a family of closed lightlike geodesics. The inner horizon is
degenerate to a null plane, with no interior, but it still reaches the
particles in the moment of collision. Moreover, it also coincides with
the event horizons, which means that there is no interior of the
wormhole, and the horizon length is zero. Again, all this agrees with
the previous results. 

To summarize, it seems that the actually rotating black hole has a much
simpler structure than the static one. This is probably a bit
surprising, because usually the situation is the other way around.
There are very elegant ways to construct rotating \emph{vacuum} black
holes \cite{SpinningAdS,ViBrill}, with no infalling matter, but, as
compared to the static ones, it is much harder to visualize them.
Usually, one only has a geometric description of their covering spaces
as subsets of anti-de~Sitter space, but unless the angular velocity
vanishes one cannot identify a primitive patch from which they can be
constructed by cutting and gluing. 

The effect of the particles seems to be that they cut away certain
pieces of these spacetimes, as described in the last section of
\cite{Matschull}, such that a visualization by cutting and gluing
becomes possible. It is also the presence of the particles, and in
particular the fact that they are massless, which makes it possible to
study a single wormhole, instead of a periodic sequence. If the
particles would not emerge from $\scri$ at some time and disappear at a
later time, then the whole spacetime would become periodic in $t$ with a
period of $\pi$. Like in the maximally extended rotating vacuum BTZ
black hole, we would then get a whole sequence of wormholes connecting a
series of external universes.

\section*{Acknowledgements}
We would like to thank Ingemar Bengtsson and Jorma Louko for many
valuable discussions, and the Albert Einstein Institute in Potsdam for
hospitality.

\begin{appendix}
\section*{Appendix}
\def\thesection{A}
\setcounter{equation}{0} 

Here we want to give some technical details which are not necessary to
understand the description of our spacetime, but which are needed to
prove some of its properties. There are basically three open problems,
namely the definition of the rotational Killing vector, the proof of
geodesic completeness, and we have to show that there is only one region
of closed timelike curves. 

\subsubsection*{The rotational Killing vector}
Consider the isometries $\riso_1$ and $\riso_2$ on anti-de~Sitter space,
as defined in \eref{riso}. On the group manifold, they can be written as
\begin{equation}
  \label{riso-grp}
  \riso_1: \quad 
     \xx \mapsto (\gam_0\uu_1)^{-1} \, \xx \, (\gam_0\vv_1), \qquad
  \riso_2: \quad 
     \xx \mapsto (\gam_0\uu_2)^{-1} \, \xx \, (\gam_0\vv_2), 
\end{equation}
with the group elements $\uu$ and $\vv$ given by \eref{p1-par} and
\eref{p2-par}. The first step is to define two Killing vectors $\kll_1$
and $\kll_2$ such that $\riso_1$ and $\riso_2$ are their flows. To do
this, we have to write the group elements appearing in \eref{riso-grp}
as exponentials. It follows from 
\begin{equation}
  \label{riso-tr}
  \Trr{\gam_0\uu_1} = \Trr{\gam_0\uu_2} 
      = - 2 \cosh(\mu/2), \quad
  \Trr{\gam_0\vv_1} = \Trr{\gam_0\vv_2} 
      = - 2 \cosh(\nu/2),
\end{equation}
that all of them are hyperbolic or, for $\mu$ or $\nu$ equal to zero,
parabolic elements of $\grp$. In any case, they can be \emph{uniquely}
written as exponentials, such that
\begin{equation}
  \label{riso-exp}
  \gam_0 \uu_1 = -\expo{\mm_1/2} ,\quad
  \gam_0 \vv_1 = -\expo{\nn_1/2} ,\quad
  \gam_0 \uu_2 = -\expo{\mm_2/2} ,\quad
  \gam_0 \vv_2 = -\expo{\nn_2/2} ,
\end{equation}
where $\mm,\nn\in\alg$ are either spacelike or lightlike Minkowski
vectors, with
\begin{equation}
  \label{riso-boost}
  \mm^2= \mu^2 \, \one, \qquad
  \nn^2= \nu^2 \, \one.
\end{equation}
Explicitly, we find that
\begin{equation}
  \label{riso-log}
  \mm_{1,2} = - \mu \, 
                \frac{\gam_0 \pm \cosh(\mu/2) \gam_1}{\sinh(\mu/2)}, 
   \qquad
  \nn_{1,2} = - \nu \, 
                \frac{\gam_0 \mp \cosh(\nu/2) \gam_1}{\sinh(\nu/2)}.
\end{equation}
The Killing vectors with the required properties
$\riso_1=\expo{\pi\kll_1}$ and $\riso_2=\expo{\pi\kll_2}$ are defined
such that
\begin{equation}
  \label{kill-grp}
  2 \pi \, \kll_1 (\xx)  = 
      \xx \, \nn_1 - \mm_1 \, \xx , \qquad
  2 \pi \, \kll_2 (\xx)  = 
      \xx \, \nn_2 - \mm_2 \, \xx , 
\end{equation}
where, as explained in \sref{math}, $\xx$ is understood as an $\grp$
valued function on anti-de~Sitter space, so that $\kll_1$ and $\kll_2$
become vector fields on $\ads_0$. 

The rotational Killing vector $\kllr$ on $\ads$ will be defined
piecewise. First we have to specify its support. Consider the flow lines
of $\kll_1$ starting off from the surface $\srf_2$. All these flow lines
end up on the surface $\srf_1$. This is because $\srf_1$ is a fixed
surface of $\iso_1$, and therefore we have
$\expo{\pi\kll_1}(\srf_2)=\riso_1(\srf_2)=
\iso_1(\rot(\srf_2))=\iso_1(\srf_1)=\srf_1$. It is also not difficult to
find out in which direction these flow leave $\srf_2$, and from which
direction they arrive at $\srf_1$. They either leave the lower face of
$\srf_2$ and arrive at the upper face of $\srf_1$ or vice verse.
However, the latter is excluded. This can be seen in \fref{fix}. If the
flow lines of $\kll_1$ start off from the upper face of $\srf_2$ and
reach $\srf_1$ from below, then they had to cross the fixed right moving
light rays $A'_2C'_1$ and $B'_2D'_1$ of $\riso_1$.

\begin{figure}[t]
  \begin{center}
    \epsfbox{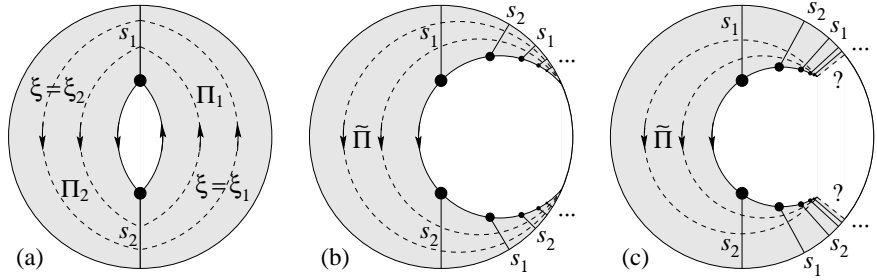}
    \caption{A schematic picture (a) of the support $\tor$ of the rotational 
      Killing vector $\kllr$. The covering space $\cov\tor$ is obtained
      by attaching infinitely many copies of $\tor_1$ and $\tor_2$ to
      each other, which are isometric to the subsets
      $\riso_2^n(\tor_2)$, $n\in\ZZ$. If $\riso_2$ has no fixed light
      rays other than those on the boundary (b), then the spacetime $\ads$
      is geodesically complete. Otherwise (c) it is incomplete.}
    \label{tor}
  \end{center}
  \hrule
\end{figure}

But this is impossible, because $\riso_1$ is the flow of $\kll_1$.
Hence, the flow lines of $\kll_1$ connect the lower face of $\srf_2$
with the upper face of $\srf_1$. Similarly, the flow lines of $\kll_2$
connect the lower face of $\srf_1$ with the upper face of $\srf_2$. Let
us denote the regions of anti-de~Sitter space filled by these flow lines 
by 
\begin{equation}
  \label{flow-torus}
  \tor_1 = \bigcup_{\tau\in[0,\pi]} \expo{\tau\kll_1}(\srf_2), \qquad
  \tor_2 = \bigcup_{\tau\in[0,\pi]} \expo{\tau\kll_2}(\srf_1).
\end{equation}
They are bounded by the cut surfaces $\srf_1$ and $\srf_2$, the boundary
$\scri$ of the cylinder, and the flow lines connecting the world lines
of the particles. We have sketched the situation in \fref{tor}(a).
Obviously, $\tor=\tor_1\cup\tor_2$ can also be regarded as a subset of
$\ads$, we only have to glue the two parts together appropriately along
the cut surfaces. Their union has the shape of a torus, with the world
lines of the particles sitting on its inner boundary. This subset of
$\ads$ is the maximal support of the rotational Killing vector $\kllr$,
which is defined to be
\begin{equation}
  \label{kllr-def}
  \kllr\big|_{\tor_1} = \kll_1  , \qquad
  \kllr\big|_{\tor_2} = \kll_2  .
\end{equation}
What we have to show is that this is indeed a Killing vector, which is
obvious everywhere except on the cut surfaces, and that its flow lines
are closed, such that $\expo{2\pi\kllr}=\id$. All this can be inferred
from the relations
\begin{equation}
  \label{riso-rel}
  \riso_1 = \iso_1 \circ \riso_2 \circ \iso_1^{-1}, \qquad
  \riso_1 = \iso_1 \circ \riso_2 \circ \iso_1^{-1},
\end{equation}
between the isometries that define the Killing vectors and the
transition function on the cut surfaces. They imply that 
\begin{equation}
  \label{riso-push}
  \iso_1^*(\kll_2)=\kll_1, \qquad 
  \iso_2^*(\kll_1)=\kll_2,
\end{equation}
where $\iso^*$ denotes the push forward of vectors induced by $\iso$.
Now, consider the vector $\kllr$ in the neighbourhood of the cut surface
$\srf_1$. We saw that the region below $\srf_1$ is part of $\tor_2$, and
that above $\srf_1$ is part of $\tor_1$. Hence, we have that
$\kllr=\kll_2$ below and $\kllr=\kll_1=\iso_1^*(\kll_2)$ above $\srf_1$.
On the other hand, $\iso_1$ is the isometry that relates an object
defined below $\srf_1$ to the same object defined above $\srf_1$, and
therefore $\kll_1$ represents the same Killing vector above $\srf_1$ as
$\kll_2$ does below $\srf_1$. Similarly, one shows that $\kllr$ is well
defined on the other cut surface $\srf_2$.
 
To prove is that the flow lines are closed and $\expo{2\pi\kllr}=\id$,
it is sufficient to show that this is the case for at least one point on
every flow line. Let us consider the points on $\srf_1$. Starting from a
point $P$ on the lower face of $\srf_1$, which is part of $\tor_2$, we
follow the flow line of $\kllr=\kll_2$ to
$\expo{\pi\kll_2}(P)=\riso_2(P)$ on the upper face of $\srf_2$. This is
identified with $\iso_2^{-1}(\riso_2(P))=\rot(P)=Q$ on the lower face of
$\srf_2$.  From there we follow the flow line of $\kllr=\kll_1$ to
$\expo{\pi\kll_1}(Q)=\riso_1(Q)$ on the upper face of $\srf_1$, which is
then identified with $\iso_1^{-1}(\riso_1(Q))=\rot(Q)=P$. Hence, we are
back to where we started after a rotation by $2\pi$.

\subsubsection*{Geodesic completeness}
To proof that our generalized Gott universe is geodesically complete, we
have to show that, given a start point and a tangent vector of a
geodesic somewhere in $\ads$, then it can always be extended arbitrarily
far into the direction of the given vector. In other words, we have to
check that the maximally extended geodesic in $\ads$ with the given
initial conditions has an infinite total length, measured in units of the
given tangent vector. The most convenient way to do this is to consider
a series of different cases, distinguished by the way in which the
geodesic intersects with the cut surfaces $\srf_1$ and $\srf_2$.

To avoid a too complicated classification, we use the following
convention regarding the conical singularities on the world lines.
Whenever a geodesic hits one of the world lines, then there are two
possible extensions. We can either take a limit from the left or from
the right. That is, we can extend the geodesic as if it had passed the
particle on the left or on the right. We regard them as two different
geodesics, and depending on which side it has passed the world line, it
either intersects with the cut surface attached to that world line or
not. Hence, it either belongs to one of the classes defined below or to
another.

The first and simplest class to be considered consists of those
geodesics in $\ads$ that, when maximally extended into one direction,
intersect only finitely many times, or not at all, with the cut
surfaces. These are certainly complete, because after the last
intersection they behave like geodesics in anti-de~Sitter space, and
this is of course geodesically complete. Another simple class of
geodesics are those that are, from the very beginning or after hitting a
world line of a particle, tangent to one of the cut surfaces. They are
necessarily lightlike or spacelike. The lightlike ones are exactly those
that span the null half planes $\srf_1$ and $\srf_2$, and these are
obviously complete.

The spacelike ones either extend to $\scri$ without leaving the cut
surfaces again, and are therefore also complete. Or, they hit the world
line and leave the cut surface, in which case we can drop a finite piece
and consider them as belonging to one of the other classes. So, actually
only those geodesics are of interest which intersect infinitely many
times with the cut surfaces. We know already that at least one such
geodesic exists, namely the closed spacelike geodesic that forms the
cusp of the inner horizon. Since this circle has a finite circumference,
it follows that its total length is infinite. What is common to all such
geodesics is that they intersect the two cut surfaces alternatingly.

This is simply because no geodesic in anti-de~Sitter space intersects a
null plane twice. But in which direction do they intersect with the cut
surfaces? It is clear that a timelike or lightlike geodesic always
intersects the cut surfaces into the same direction. That is, a future
pointing timelike or lightlike geodesic necessarily hits the lower face
and continues from the upper face. For a past pointing geodesic it is
the other way around. A spacelike geodesic can however leave the upper
face of one cut surface and hit the upper face of the other cut surface, 
or similarly leave the lower face of one cut surface and hit the other
lower face. It turns out that if it does so infinitely many times, then
it is complete. 

To see this, we have to show that there is a finite minimal length of a
spacelike geodesic that connects the two upper faces of the cut surface,
respectively the two lower faces. It should be clear from \fref{cut}
that for both the start and end point of any geodesic that connects the
upper face of $\srf_1$ with that of $\srf_2$ we must have $t\ge\pi/2$.
If we extend to two null half planes to become full planes, then they
intersect at $t=\pi/2$, and it is only the part above the intersection
where the upper faces point towards each other. Similarly, if a
spacelike geodesic connects the two lower faces, then we must have
$t\le\pi/2$ for both the start and the end point.

It is not very difficult to compute the length of a general spacelike
geodesic in anti-de~Sitter space, connecting two points represented by
the group elements $\xx,\yy\in\grp$. If $\yy=\one$ and
$\xx=\expo{d\nn}$, where $\nn\in\alg$ is a unit spacelike vector, then
the length of the geodesic is $d$, and thus $\Tr(\xx)=2\cosh d$. Using
that $\xx\mapsto\xx\yy^{-1}$ is an isometry, it follows that the length
$d$ of a geodesic connecting $\xx$ and $\yy$ is given by
$\Trr{\xx\yy^{-1}}=2\cosh d$. Using the explicit representation of the
group elements \eref{surf-sl} corresponding to points $(t,\ph)$ on the
null plane $\srf_1$, and a similar relation for points $(t',\ph')$ on
the null plane $\srf_2$, which is obtained by replacing $\gam_1$ with
$-\gam_1$ and $\gam_2$ with $-\gam_2$, one can easily show that
$d\ge2\dis$.

This holds for both $t,t'\ge\pi/2$ and $t,t'\le\pi/2$, and under the
condition that $\cos\ph\ge\tanh\dis$ and $\cos\ph'\ge\tanh\dis$, which
defines the parts of the null planes that actually form the cut
surfaces. This is quite reasonable, because $2\dis$ is the distance
between the particles in the moment of closest approach, which is
achieved at $t=\pi/2$. Note, however, that this is not the \emph{minimal} 
distance between the particles. There are points on the world line which 
are lightlikely separated, for example, but for them we have to choose
the time coordinates such that either $t<\pi/2<t'$ or vice versa. In any 
case, a spacelike geodesic that connects the two upper faces or the two
lower faces has a length of at least $2\dis$. 

If any spacelike geodesic has infinitely many such pieces, then it is
obviously complete, because it has an infinite proper length. Hence, all
what remains to be shown is that all those geodesics are complete which
intersect the two cut surfaces alternatingly and, probably after some
finite number of intersections, always into the same direction. That is,
they either always hit the surfaces from below and continue from above,
or vice versa. It is sufficient to consider the first case only. The
second is analogous, just with the time direction reversed, because
$\ads$ is invariant under a suitable chosen time inversion, combined
with a parity transformation. So, let $\crv$ be such a geodesic, and
split it into pieces $\crv_n$ connecting the upper face of one cut
surface with the lower face of the other cut surface.

To show that $\crv$ is complete, we map it isometrically
onto a geodesic in anti-de~Sitter space $\ads_0$, and show that it
converges to a point on $\scri_0$. We are not going to give the full
technical details of the proof here, which is completely straightforward
and very similar to the construction of the covering space of the
singular region in \sref{gott}. The basic idea is the following. We
start from a piece $\crv_0$ connecting the upper face of $\srf_2$ with
the lower face of $\srf_1$. This is a piece of a geodesic in $\ads$, but
we may as well regard it as a piece of a geodesic in $\ads_0$. The next
piece of the geodesic in $\ads$ is $\crv_1$, which connects the upper
face of $\srf_1$ with the lower face of $\srf_2$.

The continuation of the same geodesic in $\ads_0$ is
$\iso_1^{-1}(\crv_1)$. It starts from a point on the null half plane
$\srf_1$ and ends on $\iso_1^{-1}(\srf_2)=\riso_2^{-1}(\srf_1)$. The
next piece is $\crv_2$ in $\ads$, and it is mapped onto
$\iso_1^{-1}(\iso_2^{-1}(\crv_2))$, which is the continuation of the
geodesic in $\ads_0$. Its end point lies on the null half plane
$\iso_1^{-1}(\iso_2^{-1}(\srf_1))=\riso_2^{-2}(\srf_1)$. Continuing this
way, it is not difficult to see that we obtain a geodesic in
anti-de~Sitter space with a series of points on the null half planes
$\riso_2^{-n}(\srf_1)=\riso_2^{-n-1}(\srf_2)$. The crucial question is
what happens to them in the limit $n\to\infty$. Again, it is most
convenient to look at the conformal boundary first. 

\begin{figure}[t]
  \begin{center}
    \epsfbox{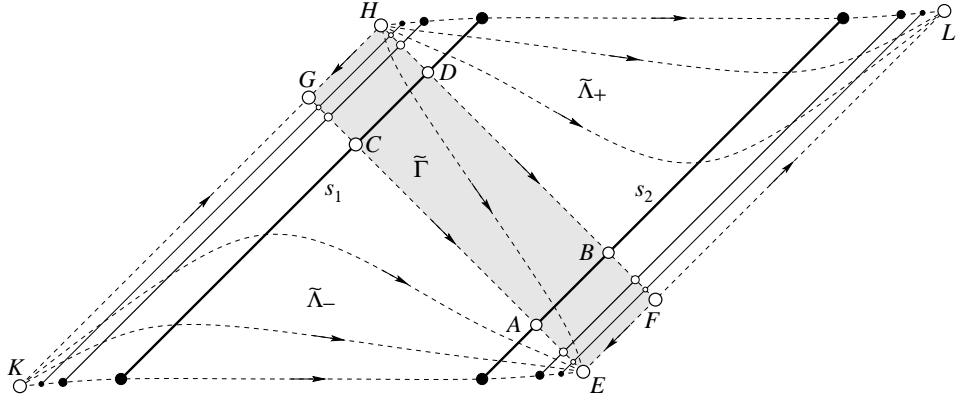}
    \caption{The flow lines of $\kll_2$ on $\scri_0$ in the region
      between the two right moving fixed light rays $KGH$ and $EFL$.
      These are the limits of the images
      $\riso_2^{-n}(\srf_1)=\riso_2^{-n-1}(\srf_2)$ for
      $n\to\pm\infty$. The total region shown here can be regarded as
      the covering space of the region in \fref{ctc}. What can be seen
      here as well is that $G$ is the apparent position of the last
      point on $\bnd_-$ and $F$ is that of the first point on $\bnd_+$.}
    \label{cov}
  \end{center}
  \hrule
\end{figure}

In \fref{cov}, we have sketched the flow lines of $\kll_2$, which is the
generating Killing vector of $\riso_2$. The region between the two bold
lines is the outer boundary of $\tor_2$, where the rotational Killing
vector is $\kllr=\kll_2$. Continuing the flow lines beyond the cuts up
to the fixed points, which we constructed in \fref{fix}, we can see what
the series of images $\riso_2^{-n}(\srf_1)=\riso_2^{-n-1}(\srf_2)$ of
the cuts looks like.  For $n\to\infty$, they converge to the fixed line
$KGH$, and for $n\to-\infty$, which we have to consider if we extend
our geodesic into the opposite direction, they converge to the fixed
line $EFL$. Note that the fixed points $K$ and $L$ could not be seen in
\fref{fix} because they are outside the time interval $0<t<\pi$.

In any case, what we see is that on the boundary of the cylinder, the
images of the cuts $\srf_1$ and $\srf_2$ converge to a pair of fixed
light rays on $\scri_0$. What does this mean for the null half planes
$\riso_2^{-n}(\srf_1)=\riso_2^{-n-1}(\srf_2)$ inside the cylinder?
Obviously, they have to converge to the null plane with origin $K$ and
destination $H$ for $n\to\infty$, respectively to the null plane with
origin $E$ and destination $L$ for $n\to-\infty$. Note that $K$ and $H$,
as well as $E$ and $L$ are antipodal points on $\scri_0$. Moreover, but
this is not of particular importance here, these are also the null
planes that form the inner horizon, which we already know to be fixed
surfaces of $\riso_2$. But now, the surfaces $\srf_1$ and $\srf_2$ are
null \emph{half} planes, so what is also important to know is what
happens to the images of the world lines
$\riso_2^{-n}(\prt_1)=\riso_2^{-n-1}(\prt_2)$.

There are three possible cases. One possibility, which is shown in
\fref{tor}(b), is that they also converge to the fixed light rays $KGH$
and $EFL$ of $\scri_0$. Then we are finished, because this immediately
implies that also the points on the geodesic under consideration
converge to $\scri_0$. We can also say that in this case the null half
planes actually disappear in the limit $n\to\infty$. Another way to
explain this is to consider the shaded region of \fref{tor}(b) as the
covering space $\cov\tor$ of the support of the rotational Killing
vector. Obviously, every geodesic in this region either reaches $\scri$,
and is thus complete, or it hits the inner boundary of $\tor$ and
continues in the complement of $\tor$ in $\ads$, which is a simply
connected subset of anti-de~Sitter space. Hence, in neither part of
$\ads$ a geodesic can get lost.

A second possible scenario is that the images
$\riso_2^{-n}(\prt_1)=\riso_2^{-n-1}(\prt_2)$ of the world lines
converge to the boundary of the cylinder, but on the opposite side of
the limiting null plane. In this case, we would say that the null half
planes become full planes in the limit. This however is excluded, for
the following reason. As can be seen in \fref{cov}, the flow lines on
which the origins and destinations of the null planes approach their
fixed points, that is, the uppermost and the lowermost flow lines shown,
are spacelike. This implies that the null planes emerging from two
points very close to each other on these lines will always overlap.
Hence, for large $n$ the surfaces $\riso_2^{-n-1}(\srf_1)$ and
$\riso_2^{-n}(\srf_1)=\riso_2^{-n-1}(\srf_2)$ would overlap. However,
$\riso_2^{-n-1}$ is an isometry, and thus $\srf_1$ and $\srf_2$ would
also overlap, and this is not the case. 

The only case that remains is the one shown in \fref{tor}(c). The world
lines converge to light rays inside the cylinder, so that in the limit
the null half planes remain half planes. In this case, the spacetime
would be geodesically incomplete, because a geodesic that approaches the
limiting null half plane has a finite length, but cannot be be extended
further. Using the interpretation of the shaded region in \fref{tor}(c)
as the covering space of $\tor$, we see that this has an additional
boundary, which is neither $\scri_0$ nor the boundary where a geodesic
passes from $\tor$ into its complement in $\ads$. As a result, a
geodesic hitting this extra boundary cannot be extended. So, the crucial
question is whether the images
$\riso_2^{-n}(\prt_1)=\riso_2^{-n-1}(\prt_2)$ of the world lines
converge inside anti-de~Sitter space or not.

Clearly, this depends on whether there are any fixed light rays of
$\riso_2$, because the limit would be such a fixed light ray. This is a
question that we already answered in the end of \sref{math}. We found
that fixed light rays exist if and only if the generating Killing vector
of $\riso_2$ is of the form \eref{kll-def} with $\mm$ and $\nn$ being
two lightlike or spacelike Minkowski vectors of the same length. Now, we
know that for the rotational Killing vector $\kllr=\kll_2$ generating
$\riso_2$, the Minkowski vectors are given by \eref{riso-log}, and these
are of the same length if and only if $\mu=\nu$. Hence, we arrive at the
final result that our spacetime is geodesically complete in all cases
except for the static one. Indeed, almost all the arguments used above
break down in the case $\dis=0$.

\subsubsection*{The region of closed timelike curves}
Finally, we want to prove some properties of the region
$\ctc\subset\ads$, which is defined to be the union of all closed
timelike curves in $\ads$. The main problem is to show that there is
only one such region, or, in other words, that $\ctc$ is connected. As a
by-product, we will also prove some statements about the extremal and
the static universes. The actual proof splits into two parts. First we
show that a region $\ctc_0\subset\ctc$ filled by a special class of
closed timelike curves is connected. Then we show that every closed
timelike curve passes through $\ctc_0$, and therefore $\ctc$ is also
connected.

It is useful to introduce an alternative set of coordinates on the cut
surfaces. Let $(t_-,t_-,\ph)$, in spherical coordinates, be a point on
the lower face of $\srf_1$, and $(t_+,t_+,\ph)$ be its counterpart on
the upper face. We then have the relation \eref{srf-iso} between $t_-$
and $t_+$, which can be written as
\begin{equation}
  \label{ctc-coor}
  \cot t_\pm = \tau \pm \jmp(\ph), \qquad
   \jmp(\ph) = \ft12 \tan\erg \, \expo{-\dis} (\cos\ph + 1) 
             + \ft12 \tan\erg \, \expo{\dis}  (\cos\ph - 1) > 0 ,
\end{equation}
where $\tau$ is some real number. Obviously, we can use $\tau$ and $\ph$
as coordinates specifying a physical point on the cut surface $\srf_1$
in $\ads$, and the same coordinates can be introduced on $\srf_2$. Their
range is such that $-\infty<\tau<\infty$ and $\tanh\dis\le\cos\ph<1$.

Now, let $(\tau,\ph)\in\srf_1$ and $(\tau',\ph')\in\srf_2$ be two points
on the cut surfaces. Can they be connected by a future pointing timelike
curve? For this to be the case, the point $(t'_-,\pi-t'_-,\pi-\ph')$ on
the lower face of $\srf_2$ must lie in the future of the point
$(t_+,t_+,\ph)$ on the upper face of $\srf_1$, with $t'_-$ and $t_+$
given by \eref{ctc-coor}. To formulate this condition as a function of
$(\tau,\ph)$ and $(\tau',\ph')$, it is most convenient to use the
conformally transformed metric, which of course has the same timelike
curves. The question is then whether the spatial distance $d$ of the two
points on the Euclidean sphere, which is given by
\begin{equation}
  \label{sphere-dis}
  \cos d = - \cos t_+ \, \cos t'_- 
           - \sin t_+ \, \sin t'_- \, \cos(\ph+\ph'),
\end{equation}
is smaller than their time distance $t'_--t_+$, for which we have
\begin{equation}
  \label{time-dis}
  \cos(t'_- - t_+) = \cos t'_- \, \cos t_+ 
                   + \sin t'_- \, \sin t_+ .
\end{equation}
Comparing the two cosines, we get the condition that
\begin{equation}
  \label{time-con-x}
  - \cot t_+ \, \cot t'_- > \ft12 ( \cos(\ph+\ph') + 1 ). 
\end{equation}
The right hand side is always positive, which implies that the signs of
the cotangents must be different. This is only possible if
$t_+<\pi/2<t_-$, because $t_+$ must be smaller than $t_-$. Using
\eref{ctc-coor}, this can be written as
\begin{equation}
  \label{time-con}
   ( \jmp(\ph) + \tau ) ( \jmp(\ph') - \tau' ) 
          > \ft12 ( \cos(\ph+\ph') + 1 ),
\end{equation}
with the additional condition that both terms in the parenthesis to the
left must be positive. This is the condition for a point $(\tau,\ph)$ on
one of the cut surfaces to be connected to $(\tau',\ph')$ on the other
cut surface by a future pointing timelike curve.

Now, consider a special class of \emph{symmetric} closed timelike
curves, and define $\ctc_0\subset\ctc$ to be the union of all these
curves. A closed timelike curve is symmetric if it intersects once with
each cut surface, at two antipodal points with the same coordinates
$(\tau,\ph)$. From \eref{time-con}, we get the condition that
\begin{equation}
  \label{sym-ctc-con}
   ( \jmp(\ph) + \tau ) ( \jmp(\ph) - \tau )
     > \ft12 ( \cos(2 \ph) + 1 )  \equivalent
    \jmp^2(\ph) - \tau^2  >  {\cos^2}\ph.
\end{equation}
Whenever this condition is satisfied, then there is a symmetric closed
timelike passing through the points with coordinates $(\tau,\ph)$ on the
two cut surfaces. Take, for example, the grand circles on the Euclidean
sphere, or the true timelike geodesics with respect to the
anti-de~Sitter metric, connecting the points on the faces of the two cut
surfaces.

Before continuing the proof that $\ctc$ is connected, let us analyze
this condition for different values of $0<\erg<\pi/2$ and $\dis\ge0$. It
is easy to show that there is no solution at all for
$\expo{-\dis}\tan\erg<1$. This is also the threshold found in
\sref{gott} for the closed lightlike curves on $\scri$ to arise. For
$\expo{-\dis}\tan\erg>1$ and $\dis>0$, we find that there is a maximal
value of $\ph$ for which $\jmp(\ph)>\cos\ph$, namely
\begin{equation}
  \cos\ph > \frac{\tan\erg \, \sinh\dis}
                 {\tan\erg \, \cosh\dis - 1 }
          > \tanh\dis.
\end{equation}
This means that there are symmetric closed timelike curves, but due to
the fact that the lower limit of $\cos\ph$ is still bigger than
$\tanh\dis$, they do not reach the world lines of the particles.
Remember that on the world lines we have that $\cos\ph=\tanh\dis$. A
special situation however arises in the case $\dis=0$. Then, the
condition says that $\cos\ph$ must be positive, but it can be
arbitrarily small. Hence, in the \emph{static} case there are closed
timelike curves in every neighbourhood of the point of collision of the
particles, which is in the new coordinates on the cut surfaces at
$\tau=0$ and $\ph=\pi/2$.

In the \emph{extremal} case, where $\expo{-\dis}\tan\erg=1$, the lower
limit for $\cos\ph$ becomes $1$, which means that the inequality above
cannot be satisfied. However, if we include the boundary at $\ph=0$,
then there is exactly one solution where \emph{equality} holds in
\eref{sym-ctc-con}, namely $\ph=0$ and $\tau=0$, and this tells us that
there is a single closed \emph{lightlike} curve on $\scri$. Finally, in
the \emph{extremal static} case, we have $\dis=0$ and $\erg=\pi/4$, and
thus $\jmp(\ph)=\cos\ph$. There is no solution for \eref{sym-ctc-con},
but when equality is allowed we find a series of solutions with
$0\le\ph\le\pi/2$ and $\tau=0$. This provides a null plane spanned by a
family of closed lightlike curves, forming the horizons, as explained in
the end of \sref{wormhole}.

Now, let us return to the generic case and the proof that the region of
closed timelike curves is connected. It is quite obvious from
\eref{ctc-coor} that whenever both $\expo{-\dis}\tan\erg$ and
$\expo{\dis}\tan\erg$ are bigger or equal to one, then $\jmp(\ph)$ is
increasing faster than $\cos\ph$ with decreasing $\ph$. In other words,
if the condition \eref{sym-ctc-con} is satisfied for some values of
$\tau$ and $\ph$, then it is also satisfied for the same $\tau$ and
smaller $\ph$. On the other hand, decreasing $\ph$ means deforming the
symmetric closed timelike curve into the direction of $\scri$, and for
$\ph=0$ it becomes a closed timelike curve on $\scri$. Hence, every
symmetric closed timelike curve can be deformed into a closed timelike
curve on $\scri$.

This proves that the region $\ctc_0$ filled with symmetric closed
timelike curves is connected, because we know that the region $\sng$ of
closed timelike curves on $\scri$ is connected. To show that this is
also true for the subset $\ctc\subset\ads$ of all closed timelike
curves, let $\crv$ be such a curve. From \sref{gott} we know that $\crv$
intersects alternatingly with the two cut surfaces. Denote by
$P_n=(\tau_n,\ph_n)$ the points of intersection, with $n$ running
cyclically around $\crv$. Since each part of $\crv$ between $P_n$ and
$P_{n+1}$ must be a timelike curve, we get the condition that, for all
$n$,
\begin{equation}
  \label{ctc-con-n}
   ( \jmp(\ph_n) + \tau_n ) ( \jmp(\ph_{n+1}) - \tau_{n+1} ) 
          > \ft12 ( \cos(\ph_n+\ph_{n+1}) + 1 )
          > \cos\ph_n \, \cos\ph_{n+1}.
\end{equation}
The right hand side of these inequalities are all positive, because
$\ph_n<\pi/2$, and therefore we can multiply them. This yields
\begin{equation}
  \label{ctc-con-prod}
   \prod_n \big( \jmp^2(\ph_n) - \tau_n^2 \big)   
          > \prod_n {\cos^2}\ph_n .
\end{equation}
The factors to the left are also positive, which implies that at least
for one of them we have
\begin{equation}
  \label{ctc-con-fac}
    \jmp^2(\ph_n) - \tau_n^2    
          > {\cos^2}\ph_n .
\end{equation}
But this is exactly the condition \eref{sym-ctc-con} for a symmetric
closed timelike curve, which means that the point $P_n$ on $\crv$ lies
in the subset $\ctc_0$ of symmetric closed timelike curves. This
completes the proof that $\ctc$ is connected.

\end{appendix}

\end{document}